\documentclass[pra,noshowkeys,preprintnumbers,amsmath,amssymb,showpacs,aps,10pt]{revtex4}

\usepackage{color}
\usepackage{graphicx}

\def\ad{a^{\dagger}}

\def\tr{\rm tr }
\usepackage{bbm,amsfonts}
\usepackage{amsmath}
\newcommand{\be}{\begin{equation}}
\newcommand{\ee}{\end{equation}}
\newcommand{\bea}{\begin{eqnarray}}
\newcommand{\eea}{\end{eqnarray}}
\def\Id{{\openone}}

    {\hspace*{\fill}$\Box$\vspace{1.5ex}\par}

\begin{document}
\title{Ground States of Fermionic lattice Hamiltonians with Permutation Symmetry}
\author{Christina V. Kraus$^{1,2,3}$, Maciej Lewenstein$^{4,5}$, and J. Ignacio Cirac$^{3}$}

\affiliation{
\mbox{$^1$Institute for Quantum Optics and Quantum Information of the Austrian Academy of Sciences, A-6020 Innsbruck, Austria}\\
\mbox{$^2$Institute for Theoretical Physics, University of Innsbruck, A-6020 Innsbruck, Austria}\\
\mbox{$^3$Max-Planck-Institute for Quantum Optics, Hans-Kopfermann-Str.\ 1, D-85748 Garching,
Germany.}\\
\mbox{$^4$ ICFO --The Institute of Photonic Sciences, Av. Carl Friedrich Gauss, num. 3, 08860 Castelldefels (Barcelona), Spain  }\\
\mbox{$^5$ ICREA -- Instituci{\'o} Catalana de Recerca i Estudis Avan\c{c}ats, Lluis Companys 23, E-08010 Barcelona, Spain} }

\begin{abstract}
We study the ground states of lattice Hamiltonians that are invariant under permutations, in the limit where the number of lattice sites, $N\to\infty$. For spin systems, these are product states, a fact that follows directly from the quantum de Finetti theorem. For fermionic systems, however, the problem is very different, since mode operators acting on different sites do not commute, but anti-commute. We construct a family of fermionic states, ${\cal F}$, from which such ground states can be easily computed. They are characterized by few parameters whose number only depends on $M$, the number of modes per lattice site. We also give an explicit construction for $M=1,2$. In the first case, ${\cal F}$ is contained in the set of Gaussian states, whereas in the second it is not. Inspired by that constructions, we build a set of fermionic variational wave functions, and apply it to the Fermi-Hubbard model in two spatial dimensions, obtaining results that go beyond the generalized Hartree-Fock theory.
\end{abstract}
\maketitle

\section{Introduction}

One of the simplest, yet very successful technique to investigate quantum many-body systems composed of bosons or spins is mean field theory. There, the use of variational wave functions of product form enables the characterization of ground state properties for certain problems. The success of this approach can be understood from different perspectives: (i) For spin systems with few--body interactions that are invariant under permutations, the exact ground state is a product state in the thermodynamic limit, where the number of spins $N \rightarrow \infty$. This can be viewed as a direct consequence of the quantum de Finetti theorem~\cite{Stroemer, Hudson, Petz, Caves, Fuchs1, Fuchs2, Koenig_Renner, Renner_de_Finetti1, Renner_de_Finetti2}: It states that any density operator of $L$ spins, $\sigma$, with a symmetric extension in the thermodynamic limit, is separable and can be written as a convex combination of product states. (ii) If we consider regular spin lattices in $d\rightarrow \infty$ dimensions with nearest-neighbor interactions and translation and rotation symmetry, monogamy of entanglement \cite{Wootters, OsborneVerstraete} implies that the ground state wave functions have to be of product type. Thus, product states provide us with the exact ground states for certain problems in the thermodynamic limit, and may thus capture the physics of other, more general situations. This explains why product states provide a valuable tool as variational wave functions, specially in high spatial dimensions.

Another very important class of quantum states constitutes the so-called Gaussian states, whose density operator can be written as a Gaussian function of creation and annihilation operators \cite{Glauber}. For bosonic systems, they provide us with a very relevant tool to describe many-body quantum systems. They capture many important phases of matter, the expectation values of physical observables can be efficiently computed, and they are the exact ground states of Hamiltonians that are quadratic in the creation and annihilation operators. For spin systems, they are intimately connected to product states in the following way: In permutationally invariant spin systems (each of them with $K$-levels), we can work in second quantization (i.e., in Fock space) and define $a_\mu$ ($\mu=1,\ldots,K$) as the operator that annihilates spins in level $\mu$. In this language,  a product state of $N$ spins is represented by $a^{\dagger N}|\Omega\rangle$, where $\Omega$ is the state with no spins (the vacuum), and $a$ is a linear combination of the $a_\mu$. Up to a normalization constant, this state is very closely related to the coherent state, $e^{\sqrt{N} a^\dagger}|\Omega\rangle$ (see \cite{Christandl, Koenig_Mitchison}). In particular, the (properly scaled) expectation value of any finite set of creation and annihilation operators coincides for those two states in the limit $N\to \infty$. Coherent states are Gaussian  \cite{Glauber}, and thus the set of (symmetric) product states can be seen as a subset of the Gaussian states in the thermodynamic limit.

Due to the success of mean field theory for bosons and spin systems, one may wonder if a similar approach exists for fermions. In fact, this concept is usually associated with the family of fermionic Gaussian states  (see eg. \cite{Bravyi} for an introduction). This family includes both Hartree-Fock and BCS states, and thus it is very widely used in many areas of physics and chemistry (see e.g. \cite{Fetter}). As their bosonic counterparts, they are exact ground states of quadratic Hamiltonians, and expectation values can be computed efficiently. The approach of taking them as variational trial wave functions is also known as generalized Hartree Fock Theory (gHFT) \cite{Lieb}, and can be extended to time-dependent problems and mixed states \cite{gHFT}, as well as to excitation spectra \cite{KrausOsborne}.

Alternatively, one may take the approaches (i) and (ii) for fermions, and investigate the family of states that exactly solves lattice systems with different symmetries, either in the thermodynamic limit or when the spatial dimension becomes infinite. One may expect that the family of states obtained by such an approach will be included in the set of Gaussian states, as in the case of spin systems. However, this is not true, at least in the case (ii) of infinite spatial dimensions. In fact, the latter approach has resulted in the development of the so-called Dynamical Mean Field Theory (DMFT)  \cite{DMFT}. DMFT has been successfully applied to many interesting situations, in particular to strongly correlated electron systems in a regime where perturbation theory breaks down. Despite its huge success, the implementation of DMFT is highly non-trivial, since it is not based on variational wave functions, thus limiting the applicability of this method. Alternatively, somewhat simpler  approaches such as \cite{DMFT_alternative} have been recently proposed, but their usefulness still remains to be investigated.

In this work we follow the first approach (i), and analyze fermionic lattice systems that are invariant under permutations, in the thermodynamic limit.  More specifically, we consider a lattice with $N$ sites, and associate to each site $M$ different fermionic modes, so that the total number of fermionic modes is $NM$. We are interested in the ground state of a Hamiltonian, $H$, that acts on pairs of lattice sites, and that is invariant under lattice permutations. As we will see, the definition of the thermodynamic limit has to be revised. Mode operators acting on different sites \emph{anti commute}, and thus some of the terms in the Hamiltonian have to be scaled with $N$ such that they give non-trivial contributions in the limit $N \rightarrow \infty$. We construct a simple family of states, ${\cal F}$, from which one can easily derive the ground state of any such Hamiltonian This is the main result of this work. Note that one can state a quantum de Finetti theorem for the fermionic problem studied here just by calculating the reduced state of such a family of states. The resulting theorem is quite different from standard de Finetti theorems (see also \cite{FermionicdeFinetti1, FermionicdeFinetti2}) due to the scaling we perform in the Hamiltonian. In contrast to the spin case, it will turn out that the set ${\cal F}$ cannot be obtained by restricting ourselves to fermionic Gaussian states, at least if the number of modes per site is sufficiently large. Furthermore, it does not contain all the ground states of lattice systems with nearest--neighbor interactions in the limit of infinite spatial dimensions [see (ii) above]. Therefore our results do not bear a clear connection neither to DMFT nor to gHFT. As we will show, our results, nevertheless, naturally lead to a simple extension of the set of fermionic Gaussian states, whose expectation values can still be computed efficiently. We will show how to use such set variationally, in order to simulate the physics of interacting fermionic lattice systems in arbitrary dimensions. As an illustration of the usefulness of this approach we will apply our method to the Fermi--Hubbard model in two spatial dimensions. The goal of the numerical part is not an exhaustive investigation of the phase diagram, but we rather aim at showing how the results go well beyond gHFT. In particular, these variational wave functions allow us to predict the appearance of pairing for repulsive interactions, a feature that is not possible within the framework of gHFT \cite{Lieb}. 

This paper is organized as follows. We start out in Section II by formulating the problem for spin systems, where the solution is trivial. Nevertheless,  guided by the proof of the quantum de Finetti theorem, we solve the problem with a technique that can be adapted to fermionic systems. In Section III, we state the problem for fermions, and show how the Hamiltonian has to be scaled depending on the local parity of the terms that appear in it. We then construct the family of states, ${\cal F}$, and explain how it can be used to address the problem we are considering. We also give an explicit construction for the case of $M=1,2$ modes per site. Finally, in Section IV, we build a family of variational wave functions inspired by our results, and apply it to the Fermi--Hubbard problem. We have also included several appendices with details or extensions of our calculations. In Appendix A, we solve the spin problem using second quantization and standard Bogoliubov-de Gennes techniques. In Appendix B, we address the problem of lattices with translation and rotation symmetries, and show how one can obtain upper and lower bounds to the energy density by using the techniques of Sections II and III. In Appendix C we give the details of the derivation of the variational method introduced in Section IV. Finally in Appendix D, we show how we establish the presence of pairing in Hubbard-like models, a result which is used in Section IV.

\section{Symmetric Hamiltonians in spin lattices}

We consider first $N$ spins, interacting according to some Hamiltonian, $H$. We assume that this Hamiltonian acts only on pairs of spins, and is invariant under permutations, i.e., $P_\pi H P_\pi=H$ for all permutations $\pi \in S_N$. One can view this scenario as a spin lattice where every spin interacts with any other in the same way. We will be interested in the set of states that minimize the energy density associated to $H$ in the limit $N\to\infty$. Such a set can be easily characterized \cite{Werner_meanfield} by making use of the quantum de Finetti theorem  \cite{Stroemer, Hudson, Petz, Caves, Fuchs1, Fuchs2, Koenig_Renner, Renner_de_Finetti1, Renner_de_Finetti2}. This theorem states that the two-spin reduced states of any state that is symmetric under permutations can be written as a convex combinations of product states, $\sigma=\mu\otimes\mu$, in the limit $N\to\infty$. Thus, the set we are interested in is composed of product states, $\rho=\mu^{\otimes N}$. We will call such set {\em Spin Symmetric Basic Set} (SSBS).

The purpose of this section is to present the derivation of this simple result in a way that later on can be adapted to the fermionic case. We will emphasize how one has to scale the different terms of the Hamiltonian with $N$, as well as how to define the energy density, such that the problem is sound. As we will see, for fermions the scaling is very subtle, so that this section will serve us as a warming up exercise. We will also introduce a simple technique to solve the problem we are interested in, and that can be easily extended to the fermionic case.

In Appendix A we give an alternative, but very useful method to solve the same problem and to obtain corrections in $1/N$ to the energy density. The method consists of using second quantization to describe the spins, and applying the standard Bogoliubov-de Gennes formalism to solve the Hamiltonian in the limit $N\gg 1$. We have included it here since it may be useful in the context of the de Finetti theorem. In Appendix B, we consider a different but related problem, namely a spin Hamiltonian with nearest-neighbor interactions, and translation and rotational invariance, in $d\to\infty$ spatial dimensions. This problem can also be solved by using product states. In fact, the solution to both this and the original problem studied in this section coincides and can be obtained by using the monogamy of entanglement \cite{Wootters, OsborneVerstraete} as well. In the case of fermions, however, the solution of those two problems is different. We will solve the first one in the next section, whereas the solution of the second problem is provided by DMFT, which requires very different methods.

Throughout this section, each spin is a $K$-level system, and we denote by $\{X^\alpha_n\}_{\alpha=1}^{K^2-1}$ a set of operators corresponding to the $n$-th spin such that, together with the identity, they form an orthonormal basis. The operators corresponding to different spins commute with each other, $[X^\alpha_n,X^\beta_m]=0$ for $n\ne m$.

\subsection{Statement of the problem}

Let us consider the terms in $H$ that contain one and two spin operators separately,
 \begin{subequations}
 \bea
 H^{(1)} &=& \sum_{n=1}^N \sum_\alpha d_\alpha X_n^\alpha = \sum_{n=1}^N h^{(1)}_n,\\
 H^{(2)} &=& \sum_{n\ne m=1}^N \sum_{\alpha,\beta} d_{\alpha,\beta} X_n^\alpha\otimes X_m^\beta = \sum_{n\ne m=1}^N h^{(2)}_{n,m},
 \eea
 \end{subequations}
where the coefficients $d$ are such that the corresponding operators are Hermitian. It is clear that $||H^{(1)}||_\infty = N ||h^{(1)}||_\infty$, whereas $c_2 N(N-1)\le ||H^{(2)}||_\infty\le N(N-1)||h^{(2)}||_\infty$,
 \be
 c_2 = \max_{||\varphi||_2=1} |\langle \varphi,\varphi|h^{(2)}|\varphi,\varphi\rangle| \ne 0.
 \ee
Thus, if we wish that $H^{(1)}$ and $H^{(1)}$ are both relevant when determining the ground state energy density, $E_0$, in the limit $N\to\infty$ we will have to take
 \be
 H = (N-1) H^{(1)} + H^{(2)} = \sum_{n\ne m=1}^N h_{n,m},
 \ee
where $h_{n,m}= h_{n,m}^{(2)} + (h_n^{(1)}+h_m^{(1)})/2$, and
 \be
 \label{E0a}
 E_0 = \lim_{N\to\infty} \frac{1}{N^2} \min_{||\rho||_1=1} {\rm tr}(H\rho),
 \ee
where $\rho=\rho^\dagger\ge 0$.

As mentioned above, the ground state energy density (\ref{E0a}) can be easily obtained by using the quantum de Finetti theorem \cite{Stroemer, Hudson, Petz, Caves, Fuchs1, Fuchs2, Koenig_Renner, Renner_de_Finetti1, Renner_de_Finetti2}. The states $\rho$ that minimize the energy density (\ref{E0a}) are product states, which are widely used in mean field theories. Thus, as it is well known, mean field theory is exact when dealing with lattices with permutation symmetry in the thermodynamic limit. Since the main goal of this paper is to obtain a similar results for fermions, we will re-derive the above result using a technique that can be extended to that problem. This technique is based on the proof of the quantum de Finetti theorem \cite{Christandl}. Following \cite{Christandl}, it is more convenient to work with a specific purification of $\rho$,
 \be
 \label{Psi0}
 |\Psi\rangle = (\sqrt{\rho}\otimes \Id) |\Phi^+\rangle^{\otimes N},
 \ee
where
 \be
 |\Phi^+\rangle = \sum_{k=1}^K |k,k\rangle,
 \ee
is a state defined on each lattice with two spins per site, the original  and the purifying one. The operator $\sqrt{\rho}$ in Eq. (\ref{Psi0}) acts only on the original spins. The state $|\Psi\rangle$ is an element of ${\cal H}^{\rm sym}_{N,K^2}$, the Hilbert space of $N$ spin systems with $K^2$ levels, that is invariant under lattice permutations (that permutes the original as well as the purifying spins). Furthermore,
 \be
 \rho = {\rm tr}_p \left( |\Psi\rangle\langle \Psi|\right),
 \ee
where the trace is over the purifying spins. Thus, $|\Psi\rangle$ is indeed a purification of $\rho$. Thus, we can write
 \be
 \label{E0d}
 E_0 = \lim_{N\to \infty} \frac{1}{N^2} \min_{|\Psi\rangle \in {\cal H}^{\rm sym}_{N,K^2}}\langle\Psi|H|\Psi\rangle = \lim_{N\to \infty}  \min_{|\Psi\rangle\in {\cal H}^{\rm sym}_{N,K^2}}\langle\Psi|h_{1,2}|\Psi\rangle.
 \ee
Our goal is to find the family of states, ${\cal F}\subset {\cal H}^{\rm sym}_{N,K^2}$, that attains the minimum in this definition. 
\subsection{Spin Symmetric Basic States (SSBS)}

Let us now consider a state $|\Psi\rangle\in {\cal H}^{\rm sym}_{N,K'}$, were $K'=K^2$. We can easily identify a complete set of states that span that space, namely
$\{|\phi\rangle^{\otimes N}\}$. Thus, we can express
 \be
 |\Psi\rangle = \int d \mu_{\phi} f(\phi)|\phi\rangle^{\otimes N} ,
 \ee
where $f$ is a complex function and $d \mu_{\phi}$ a measure in ${\cal H}_{K'}$. The function $f$ can be chosen to be smooth. In fact, expressing the
coefficients of $\phi$ in an orthonormal (spin) basis, $\{|n\rangle\}_{n=1}^{K'}$, as
 \begin{subequations}
 \bea
 c_n &=&\cos(\theta_1)\ldots\cos(\theta_{n-1})\sin(\theta_{n})e^{i\varphi_n}, \quad n=1,\ldots,K'-1,\\
 c_{K'} &=& \cos(\theta_1)\ldots\cos(\theta_{K'-1})\cos(\theta_{K'}),
 \eea
 \end{subequations}
and choosing the standard measure for the solid angle, we see that the highest Fourier components of $f$ in the expansion in terms of $\theta_n$ and $\varphi_n$ do not contribute to the integral. We have
 \be
 \langle\Psi|h_{1,2}|\Psi\rangle = \int d\mu_{\phi} d\mu_{\phi'} \bar f(\phi) f(\phi') \langle\phi,\phi|h_{1,2}|\phi',\phi'\rangle \; \langle \phi|\phi'\rangle^{N-2}.
 \ee
Since the overlap of the two normalized wave functions, $|\phi\rangle$ and $|\phi'\rangle$, is less or equal one, we can write $\langle \phi'| \phi\rangle^{N-2} \approx (e^{-\theta^2/2})^{(N-2)}$, where $\theta$ is the angle between $|\phi\rangle$ and $|\phi'\rangle$. Hence,
 \be
 \lim_{N\to\infty}\langle \phi|\phi'\rangle^{N-2} \propto \delta(\phi-\phi').
 \ee
This implies that in this limit, we will attain the minimum in (\ref{E0d}) if we simply take
 \be
 \label{spinMF}
 |\Psi\rangle = |\phi\rangle^{\otimes N},
 \ee
and minimize with respect to $\phi$. Thus, the states (\ref{spinMF}) form the SSBS. They are product states, and if we calculate the reduced state by tracing the purifying spins, we will obtain $\sigma=\mu^{\otimes N}$. As anticipated, states of the form (\ref{spinMF}) are those used in mean field theories. We also recover the result following from the quantum de Finetti theorem, namely that in order to determine $E_0$ we just have to solve a two-spin problem, and minimize the energy density with respect to density operators of the form $\sigma=\mu^{\otimes 2}$. As a side remark, note that we can take an asymptotic expansion of $\langle\phi|\phi'\rangle^{N-2}$ in terms of $1/N$ (which gives corrections to the delta function involving its derivatives), and in this way obtain corrections to the de Finetti theorem \cite{Koenig_Renner, Christandl}. In Appendix A we show how such corrections can be obtained by using standard techniques of statistical mechanics based on second quantization. In Appendix B we show how the solution of the present problem can be used to obtain upper and lower bounds of the energy in the case of a lattice systems with nearest-neighbor interactions in the limit where the number of spatial dimensions $d\to\infty$. We also show that for the case of spins, the product state (\ref{spinMF}) also attains the minimal energy density.

\section{Symmetric Hamiltonians in fermionic lattices}

In the previous section we learned a strategy for the characterization of the SSBS: First, we express the density operators that are invariant under permutations in terms of purifications in ${\cal H}^{\rm sym}_{N,K^2}$; then, we find a complete set of vectors, ${\cal F}$, characterized by some parameters, whose number only depends on $K$, but not on $N$. This set has the property that if we write the purification as a linear combination of its elements and determine the expectation value of a two-site operator, the cross terms vanish in the thermodynamic limit. Thus, we can just take the states in this set (and not superpositions thereof) when we minimize the energy density in that limit. Furthermore, the minimization just requires solving a two-spin problem.

In this section we will apply an analogous strategy for fermionic systems.
We consider now a fermionic lattice system with $N$ sites and $M$ modes per site. We are interested in the ground state of a Hamiltonian, $H$, acting on pairs of sites, that is invariant under site permutation, and in the limit $N\to\infty$. We will first show that every symmetric fermionic quantum state has a symmetric purification with $M' = 2M$ modes per site. Then, we construct the set of states ${\cal F}\subset {\cal H}^{\rm sym}_{N,M'}$, the Fock space of symmetric states under permutations with $M'$ modes per site, with the same properties as in the spin case. We call the states in the set ${\cal F}$ \emph{Fermionic Symmetric Basic States} (FSBS). Sometimes, we will write ${\rm FSBS}_M$ to indicate the number of modes per site. Finally, we will explicitly construct the FSBS for $M=1,2$ (i.e. $M'=2,4$). In the first case, we obtain that the FSBS are contained in the set of Gaussian states, whereas in the latter we obtain a class of states that goes beyond. Inspired by this result, we will introduce a new family of states that extends the Gaussian variational ans\"atze in the next section, and we will apply it to the Hubbard model.

In this and the following sections, we will use the language of second quantization. We denote the annihilation operators acting on mode $\mu$ at site $n$ by $a_{n,\mu}$, where $n=1,\ldots,N$, and $\mu=1,\ldots,M<\infty$. We will call the operators $a_{n,\mu}$ and $a_{n,\mu}^\dagger$ mode operators. As opposed to the spin case, two mode operators corresponding to two different sites \emph{anti commute}, i.e. $\{a_{n,\mu},a_{m,\nu}\}=\{a_{n,\mu},a_{m,\nu}^\dagger\}=0$ if $m\ne n$ or $\mu\ne \nu$ (or both). Besides that, $\{a_{n,\mu},a_{n,\mu}^\dagger\}=1$. This has important consequences in the way we have to scale the different terms of the Hamiltonian in the limit $N\to \infty$. Thus, we will start this section by defining the problem we want to solve, and argue about the corresponding scaling.

\subsection{Statement of the problem}

We consider a Hamiltonian, $H$, invariant under permutations, and with terms involving two lattice sites at most. In order to comply with the fermion parity superselection rule, each term in $H$ must contain an even number of mode operators. Thus, we can consider three kinds of terms in the Hamiltonian: (i) $H^{(1)}$ contains terms acting on single sites only, and which are composed of an even number of mode operators; (ii) $H^{(2)}_{\rm ee}$ contains terms acting on two different sites, but with an even number of mode operators on each of the two sites where they act; (iii) $H^{(2)}_{\rm oo}$ contains terms acting on two different sites, but with an odd number of mode operators on each of the two sites where they act. For instance, for $M=2$ we can have
 \begin{subequations}
 \label{Hubbardgeneral}
 \bea
 H^{(1)} &=& \mu \sum_{n=1}^N \sum_{\nu=1}^2 a^\dagger_{n,\nu} a_{n,\nu} + U \sum_{n=1}^N a^\dagger_{n,1}a^\dagger_{n,2} a_{n,2}a_{n,1},\\
 H^{(2)}_{\rm ee} &=& V \sum_{n,m=1}^N \sum_{\nu,\nu'=1}^2 a^\dagger_{n,\nu} a^\dagger_{m,\nu'}a_{m,\nu'}a_{n,\nu},\\
 H^{(2)}_{\rm oo} &=& - t \sum_{n,m=1}^N \sum_{\nu=1}^2 a^\dagger_{n,\nu} a_{m,\nu},
 \eea
 \end{subequations}

As in the previous section, let us compute the scaling of the norms for each of the terms. First of all, since $H^{(1)}$ and $H^{(2)}_{\rm ee}$ only contain an even number of mode operators for each site on which they act, they can be mapped to spins, and thus scale as $N$ and $N^2$, as before. However, the term $H^{(2)}_{\rm oo}$ does not scale as $N^2$, but rather as $N$. This can be directly seen in the example given above. We can write $H^{(2)}_{\rm oo}=-t N (A^\dagger_1 A_1 + A^\dagger_2 A_2)$, where
 \be
 A_\mu = \frac{1}{\sqrt{N}}\sum_{n=1}^N a_{n,\mu}
 \ee
fulfill standard anticommutation relations. The operators $A_\mu^\dagger A_\mu$ have eigenvalues 0 and 1, so that $||H^{(2)}_{\rm oo}||_\infty =t N$.

In general, we can write
 \be
 H^{(2)}_{\rm oo}= \sum_{\alpha} X^\alpha Y^\alpha,
 \ee
where
 \be
 X^\alpha = \sum_{n} X^\alpha_n, \quad  Y^\alpha = \sum_{n} Y^\alpha_n,
 \ee
and $X^\alpha_n$ and $Y^\alpha_n$ are products of an odd number of mode operators. Now,
 \be
 || H^{(2)}_{\rm oo}||_\infty \le \sum_{\alpha} ||X^\alpha Y^\alpha||_\infty
 \le c(M) \max_\alpha \sqrt{ ||X^{\alpha\dagger} X^\alpha||_\infty
 || Y^{\alpha\dagger} Y^\alpha||_\infty },
 \ee
where $c(M)$ is a constant that only depends on the number of modes, $M$.
Further,
 \be
 ||X^{\alpha\dagger} X^\alpha||_\infty
 \le ||\{X^{\alpha\dagger}, X^\alpha\}||_\infty
 \le 2 \sum_{n=1}^N || X^{\alpha\dagger}_n X^\alpha_n||_\infty
 = 2N ||X^{\alpha\dagger}_1X^\alpha_1||_\infty,
 \ee
where we have used that for all $n\ne m$
 \be
 \{X^\alpha_n,X^\beta_m\}=\{Y^\alpha_n,Y^\beta_m\}=0.
 \ee
Using analogous arguments for $Y^\alpha$, we obtain
 \be
 ||H^{(2)}_{\rm oo}||_\infty \le N 2c(M) ||X^{\alpha\dagger}_1X^\alpha_1||_\infty^{1/2}
 ||Y^{\alpha\dagger}_1Y^\alpha_1||_\infty^{1/2}.
 \ee

Let us draw some consequences from that fact. If we do not scale $H^{(2)}_{\rm oo}$ (i.e. if we do not multiply by $N$), it will not give any contribution to the energy density. That is, if we take (\ref{E0a}) with
 \be
 \label{Hambad}
 H = \sum_{n,m} h_{n,m} = N H^{(1)} + H^{(2)}_{\rm ee}+ H^{(2)}_{\rm oo},
 \ee
then $H^{(2)}_{\rm oo}$ will not contribute in the limit $N\to \infty$, and thus can be omitted. Since the remaining terms $N H^{(1)} + H^{(2)}_{\rm ee}$ conserve parity locally (in each site), we can map the fermionic Hamiltonian into a local spin Hamiltonian (using the appropriate Jordan-Wigner transformation): that is, the problem of minimizing the energy density reduces
to the spin problem analyzed in the previous section. We can thus take product states in order to solve the minimization problem. Analogously, in the case of a lattice with translation and rotation symmetry in the limit $d\to \infty$ we obtain that the energy can be minimized by taking product states (see Appendix B).

However, the above result is not very useful if we want to develop techniques in many-body problems involving fermions. The reason is that in those problems, the interesting regimes occur when the terms $H^{(2)}_{\rm oo}$ contribute to the problem. For instance, in the example considered above (\ref{Hubbardgeneral}), $H^{(2)}_{\rm oo}$ describes the kinetic energy of fermions moving between different sites. The richest behavior occurs when the kinetic energy is comparable to the interaction energies contained in $H^{(1)}$ and $H^{(2)}_{\rm ee}$. This implies that instead of (\ref{Hambad}), we have to consider Hamiltonians of the form
 \be
 \label{HFermions}
 H = N H^{(1)} + H^{(2)}_{\rm ee} + N H^{(2)}_{\rm oo}
 \ee
In the following we will concentrate on this case, i.e. we will find the states that minimize the energy density (\ref{E0a}) with $H$ given in Eq.~(\ref{HFermions}). Note that we can define the same problem in the lattice in $d$ dimensions, and in the limit where  $d\to \infty$. In fact, this problem is solved by DMFT. There is, however, no a priori reason why the solution we find bears any relation to DMFT.

Let us now argue that, as in the spin case, every symmetric fermionic mixed state has a symmetric purification. To this end, let $\rho\in {\cal S}_{N,M}$, be the (convex) set of density operators that are invariant under permutations and that commute with the fermion parity operator. This last property is a consequence of the superselection rule (or, equivalent, that the Hamiltonians we are considering also possess that property). We double the number of modes per site and denote by $b_{n,\mu}$ the corresponding mode operators. We define a set of states
 \be
 |\{m_{n,\mu}\}\rangle = \prod_{n,\mu} \left[ a^\dagger_{n,\mu} b^\dagger_{n,\mu}\right]^{m_{n,\mu}} |\Omega\rangle,
 \ee
where $|\Omega\rangle$ is the vacuum and  $m_{n,\mu}=0,1$.  Now, consider the state $|\Psi\rangle$ corresponding to $M'=2M$ modes, and
 \be
 |\Psi\rangle = \sqrt{\rho} \sum_{\{m_{n,\mu}\}} |\{m_{n,\mu}\}\rangle.
 \ee
For any operator $O$ depending on the fermionic operators $a_{n,\mu}$ and commuting with the fermion parity operator, we have
 \be
 \langle \Psi|O|\Psi\rangle = \sum_{\{m_{n,\mu}\}} \langle \Omega| \left(\prod_{n,\mu} \left[ a_{n,\mu} \right]^{m_{n,\mu}}\right) \sqrt{\rho}O\sqrt{\rho}
 \left(\prod_{n,\mu} \left[ a^\dagger_{n,\mu} \right]^{m_{n,\mu}}\right)| \Omega\rangle={\rm tr}(O\rho).
 \ee
Furthermore, by construction,
 \be
 \label{PureSymmetric}
 P_\pi|\Psi\rangle=|\Psi\rangle,
 \ee
for all $\pi\in S_N$, i.e. $|\Psi\rangle\in {\cal H}^{\rm sym}_{N,2M}$ is a symmetric purification of $\rho$.

Our goal is to find the FSBS, ${\cal F} \subset {\cal H}^{\rm sym}_{N,2M}$, such that 
 \be
 \label{E_0FSBS}
 E_0 = \lim_{N\to\infty} \min_{|\Psi\rangle\in{\cal F}} \frac{\langle \Psi|H|\Psi\rangle}{N^2 \langle \Psi|\Psi\rangle},
 \ee
where $H$ is given in (\ref{HFermions}). Furthermore, we want to show that this quantity can be obtained by solving a few-site problem.

\subsection{Fermionic Symmetric Basic States (FSBS)}

\begin{figure}[t]
 \begin{center}
 \includegraphics[width = 0.7 \columnwidth]{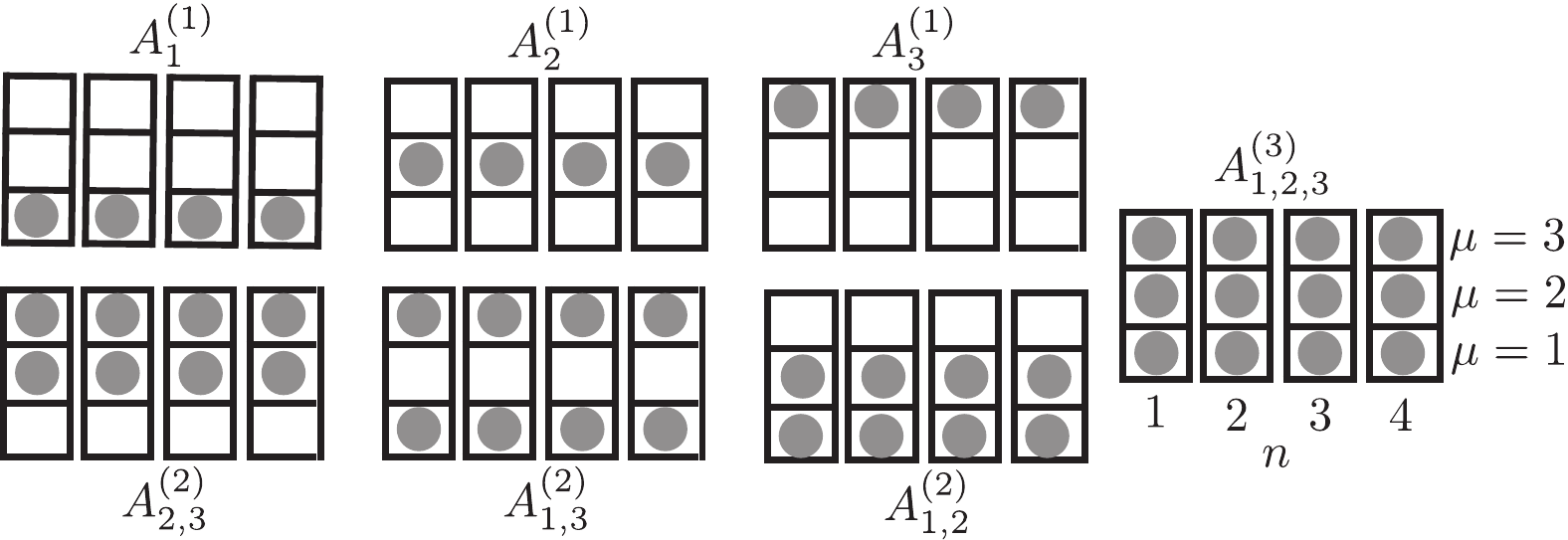}
 \end{center}
\caption{Example of the operators $\bar A_{\vec \mu}^{(k)}$, Eq.~(\ref{barA}), for $N=4$ and $M=3$. The grey balls represent the operators $\ad_{n,\mu}$ on site $n$ in mode $\mu$.\label{fig:OperatorsA}}
 \end{figure}

In order to determine the FSBS, we will closely follow the procedure for spins. We start by finding a complete set in ${\cal H}^{\rm sym}_{N,M'}$, where $M'=2M$ is always even, since we have shown above that all symmetric states have a symmetric purification with an even number of modes per site. Thus, from now on, we will drop the prime and simply write $M$. Having that set at hand we will build later on the FSBS. We define the averaged operators
 \be
 \label{barA}
 \bar A_{\vec\mu}^{(k)}:=\bar A_{\mu_1,\ldots,\mu_k}^{(k)} = \sum_{n=1}^N \prod_{l=1}^k a_{n,\mu_l},
 \ee
where $\mu_1<\mu_2<\ldots<\mu_k \le M$ are integers denoting different modes (see Fig. \ref{fig:OperatorsA}). In order to simplify the notation, we will simply write $\bar A_{\mu}^{(k)}$ in the following, and we will even omit the script $k$ whenever it is clear from the context. As we show now, any symmetric pure state can be written as a linear combination of states
 \be
 \label{SymmFock}
 \prod_\mu \left(\bar A_\mu^\dagger\right)^{n_\mu} |\Omega\rangle,
 \ee
where $n_\mu$ are integers. All those states are symmetric, given that $P_\pi  \bar A_\mu^\dagger P_\pi =\bar A_\mu^\dagger$ for all $\pi\in S_N$, and thus they will form a complete set in ${\cal H}^{\rm sym}_{N,M}$, as we will see now.

Let us consider a state $|\Psi\rangle\in {\cal H}^{\rm sym}_{N,M}$, which we write as
 \be
 |\Psi\rangle = \sum_{\vec n} X_{\vec n} |\Omega\rangle,
 \ee
where $X_{\vec n}$ is a product of creation operators,
 \be
 \label{SampleState}
 X_{\vec n} = \prod_{n_1<n_2<\ldots <n_k} O_{n_k}[k] \ldots O_{n_2}[2] O_{n_1}[1] |\Omega\rangle.
 \ee
Here, we have collected all the creation operators acting on a site $n_j$ that contain at least one fermion, denoted their product by $O_{n_j}[j]$, and sorted them with increasing site label $j$. Since $|\Psi\rangle$ is a symmetric state, it follows that
 \be
 \label{PermTrick}
 |\Psi\rangle \propto \sum_{\vec n} \sum_{\pi\in R\subset S_N} P_\pi X_{\vec n} P_\pi |\Omega\rangle,
 \ee
for any subset $R \subset S_N$ of permutations. We show now that by using the appropriate sequences of permutations in Eq. (\ref{PermTrick}), we can always obtain a linear combination of states in the desired form (\ref{SymmFock}). Consider first $O_{n_1}[1]$, and take $R$ to be the set of transpositions of $(n_1,n)$, where $n \notin \{n_2, \ldots, n_k\}$. Then, $P_{\pi\in R}$ commutes with all the $O_{n_j}[j]$, except for $O_{n_1}[1]$, and we can thus write
 \begin{eqnarray*}
 \sum_{\pi\in R\subset S_N} P_\pi X_{\vec n} P_\pi  =  \prod_{n_2<n_2<\ldots <n_k} O_{n_k}[k] \ldots O_{n_2}[2] \sum_{\pi\in R\subset S_n}P_\pi  O_{n_1}[1] P_\pi.
 \end{eqnarray*}
Further, since $R$ is the set of transpositions of $(n_1,n)$, where $n \notin \{n_2, \ldots, n_k\}$, we have
 \begin{align*}
 \sum_{\pi\in R\subset S_N}P_\pi O_{n_1}[1] P_\pi &= \bar A[1]^\dagger - \sum_{l=2}^k O_{n_l}[1],\\
 \bar A[1]^\dagger &= \sum_{n=1}^N O_n[1],
 \end{align*}
and $O_{n_l}[1]$ is the operator obtained from $O_{n_1}[1]$ by permuting the sites $n_1$ and $n_l$. Thus,
 \begin{align}
 \label{eq:decompose}
 \sum_{\pi\in R\subset S_N} P_\pi X_{\vec n} P_\pi |\Omega\rangle =& \prod_{n_2<n_2<\ldots <n_k} O_{n_k}[k] \ldots O_{n_2}[2] \bar A[1]^\dagger|\Omega\rangle\nonumber\\
 & - \sum_{l=2}^k \prod_{n_2<n_3<\ldots <n_k} O_{n_k}[k] \ldots O_{n_2}[2] O_{n_l}[1]|\Omega\rangle.
 \end{align}
Here, the second term is a sum of terms of the original form (\ref{SampleState}), but in which the operators acting on site $n_1$ are no longer present. Thus, we can repeat the procedure with such terms, until we are left with terms that contain one operator $\bar A$, as in the first part of Eq.\ (\ref{eq:decompose}). So, let us consider the first term in (\ref{eq:decompose}), which looks like (\ref{SampleState}), but with $O_{n_1}[1]$ replaced by $\bar A[1]^\dagger$. We repeat the steps that we have done for the operator $O_{n_1}[1]$ but now for the operator $O_{n_2}[2]$. Since $\bar A[1]^\dagger$ is symmetric under permutations, we arrive in a second step at a similar expression as in Eq. (\ref{eq:decompose}), but with an additional term $\bar A[2]^\dagger$, symbolically
 \[
 \prod_{n_2<n_2<\ldots <n_k} O_{n_k}[k] \ldots O_{n_2}[2] \bar A[1]^\dagger \mapsto \prod_{n_3<n_2<\ldots <n_k} O_{n_k}[k] \ldots O_{n_3}[3] \bar A[2]^\dagger \bar A[1]^\dagger.
 \]
Hence, repeating the procedure for  $O_{n_3}[3], \ldots, O_{n_k}[k]$, we can decompose the state $|\Psi\rangle$ as in (\ref{SymmFock}).

Now, let us consider the set of operators (\ref{barA}). It is clear that $\bar A^\dagger_\mu \bar A^\dagger_\nu = \epsilon_{\mu,\nu} \bar A^\dagger_\nu \bar A^\dagger_\mu$, where $\epsilon_{\mu,\nu}=\pm 1$. This implies that when we write a complete set of states we can choose any ordering of the operators. On the one hand, if $k$ is odd, $(\bar A^{(k)\dagger}_\mu)^2=0$, so that these operators can appear at most once in the states
(\ref{SymmFock}). On the other, if $k$ is even, then several powers may appear. Thus, we can write a complete set of states as
 \be
 \label{Symbasis0}
 \prod_{k\, {\rm odd}}\prod_\mu \left(\bar A^{(k)\dagger}_{\mu}\right)^{n_{k,\mu}}
 \prod_{k' {\rm even}}\prod_{\mu'} \left(\bar A^{(k')\dagger}_{\mu'}\right)^{m_{k',\mu'}} |\Omega\rangle,
 \ee
where $n_{k,\mu}=0,1$ and $m_{k',\mu'}$ is an integer. Note that when we expand the products in terms of creation operators, we can omit the terms where two sites coincide. The reason is that all those terms are already included in the operators with $k'$ odd, and this may simplify the evaluation of expectation values. Nevertheless, we will not consider that in the following. Since we are considering a complete set, we can alternative choose
 \be
 \label{Symbasis}
  |\varphi_{\vec n,\vec \alpha}\rangle=\prod_{k\, {\rm odd}}\prod_\mu \left(\bar A^{(k)\dagger}_{\mu}\right)^{n_{k,\mu}}
 \prod_{k' {\rm even}}\prod_\nu e^{\alpha_{k',\nu} \bar A^{(k')\dagger}_{\nu}} |\Omega\rangle.
 \ee
Those states are completely parametrized in terms of $n_{k,\mu}=0,1$ and $\alpha_{k',\nu}\in \mathbb{C}$. It is an important fact that the number of parameters depends only on the number of modes per site, $M$, but not on the number of lattice sites $N$. For instance, the number of indices $n$ is given by
 \be
 r(M)= \sum_{k=1}^{M/2} \left(\begin{array}{c} M\\ 2k-1\end{array}\right).
 \ee
Note that we can consider $\vec n$ as a bit string (taking values 0 and 1).

Any state in ${\cal H}^{\rm sym}_{N,M}$ can be written as a linear combination of states of the form (\ref{Symbasis}). This involves sums over the bit string $\vec n$, and integrals over the $\alpha$'s, i.e.,
 \be
 |\Psi\rangle = \sum_{\vec n} \int d\mu_{\vec \alpha} f_{\vec n}(\vec \alpha)
 |\varphi_{\vec n,\vec \alpha}\rangle.
 \ee

In order to obtain the energy density (\ref{E_0FSBS}), we have to determine expectation values of operators $O$ acting on up to two-sites, say 1 and 2, i.e. $O_{1,2}$. We will have to determine integrals over $\vec\alpha$ and $\vec\beta$ to calculate
 \be
 \langle\varphi_{\vec m,\vec \beta} |O_{1,2}|\varphi_{\vec n,\vec \alpha}\rangle.
 \ee
If we expand all operators $\bar A^{(k)}$ with $k$-odd in Eq. (\ref{Symbasis}) in terms of the lattice sites (\ref{barA}) for the states $|\varphi_{\vec n,\vec \alpha}\rangle$ and $|\varphi_{\vec m,\vec \beta}\rangle$ we will have a sum of terms of the form
 \be
 \label{aux6}
 \langle \eta(\vec\beta)| O_{m_1} \ldots O_{m_{r(M)}} O_{1,2} O_{n_1}\ldots O_{n_{r(M)}} |\eta(\vec\alpha)\rangle,
 \ee
where $O_n$ is some operator acting on site $n$ only, and
 \be
 |\eta(\vec\alpha)\rangle= \prod_{k\, {\rm even}}\prod_\nu e^{\alpha_{k,\nu} A^{(k)\dagger}_{\nu}} |\Omega\rangle.
 \ee
Now, we expand the operators $\bar A^{(k)}_{\mu}$ with $k$-even (\ref{barA}) in this expression and replace them in (\ref{aux6}). They are sums of an even number of creation (annihilation) operators, so that they commute among each other. Furthermore, we can always find $N-g_M$ sites such that no operator $O$ in (\ref{aux6}) acts on them, where $g_M$ does not depend on $N$. Thus, for each $k$-even, and each value of $m_i,n_j$ in (\ref{aux6}), we can take as a common factor
 \be
 \left(\langle \Omega| \left[\prod_{k' {\rm even}}\prod_\nu e^{\bar\beta_{k',\nu} a_{n,\nu_1}^\dagger\ldots a_{n,\nu_{k'}}^\dagger} \right]
 \left[\prod_{k\, {\rm even}}\prod_\mu e^{\alpha_{k,\mu} a_{n,\mu_1}^\dagger\ldots a_{n,\mu_k}^\dagger} \right]|\Omega\rangle\right)^{N-g_M}.
 \ee
This expression can be understood as an overlap $\langle \tau(\vec\beta)|\tau(\vec\alpha)\rangle^{N-g_M}$ of two wave functions. When we divide by their normalization, we see that it tends to something proportional to a $\delta$ function of $\vec \alpha$ and $\vec \beta$ in the thermodynamic limit. Thus, as in the case of spin systems, we can fix the values of $\alpha$'s in the thermodynamic limit  and do not need to consider superpositions thereof. That is, the vectors
 \be
 \label{FSBS}
 |\Psi\left(\{c_{\vec n}\},\vec \alpha\right)\rangle = \sum_{\vec n} c_{\vec n}  |\varphi_{\vec n,\vec \alpha}\rangle
 \ee
of $2M$ modes in the limit $N\to\infty$ form a FSBS.

Some remarks are in order. First, as in the case of spins, the set of states (\ref{FSBS}) is characterized by a finite set of complex parameters ($c_{\vec n}$ and $\vec \alpha$ for all bit strings $\vec n$), whose number does not depend on $N$. Second, we can restrict the allowed $\vec n$ by imposing the fermion parity superselection rule, i.e.  $\sum_{ k\,{\rm odd},\mu} k\, n_{k,\mu}$ to be either even or odd. Third, in order to determine the energy density, one should consider a minimization for a fixed $N$, and then take $N\to \infty$. In the next subsections we analyze the case of $M=1,2$ (i.e. $M'=2,4$), and show how one can take the limit $N\to\infty$ first, so that the problem is highly simplified. In fact, we will show that the determination of the energy density basically reduces to a three-site problem.

\subsection{Example: $M=1$}

We consider the simplest case of one mode per site, $M=1$, with annihilation operators $a_{n,1}$. The most general, symmetric Hamiltonian with two-site interactions, and that respects the parity superselection rule is of the form
 \be
 \label{Hama}
 \hat H=\frac{1}{N^2} H = -\frac{t}{N} \sum_{n,m} a_{n,1}^\dagger a_{m,1} + \frac{U}{N^2} \sum_{n,m} a_{n,1}^\dagger a_{n,1} a_{m,1}^\dagger a_{m,1}+ \frac{\mu}{N} \sum_{n} a_{n,1}^\dagger a_{n,1},
 \ee
where we have added the scaling factors as discussed in the previous subsections. Note that terms with an odd number of creation/annihilation operators are forbidden by parity, and that terms like
 \[
 \sum_{n,m} a^\dagger_n a^\dagger_m + h.c.\propto \bar A_1^{\dagger 2}+ h.c.=0,
 \]
vanish identically, where 
 \be
 \bar A_1= \frac{1}{\sqrt{N}}\sum_n a_{n,1}.
 \ee 

Determining the minimum energy of $\hat H$ for all values of $t,U$, and $\mu$ is rather trivial. The number operator,
 \be
 \hat L=\sum_{n} a_{n,1}^\dagger a_{n,1},
 \ee
commutes with $\hat H$, so that the terms in (\ref{Hama}) with $U$ and $\mu$ (which are proportional to $\hat L^2$ and $\hat L$, respectively) can be considered independently. Furthermore, the first term is proportional to $\bar A_1^\dagger \bar A_1$, which has eigenvalues $1$ and $0$. Thus, if we denote by $\rho_0=L/N$ the density, where $L$ is the number of particles, we have
 \be
 \label{E0tU}
 E_0= - t x + U \rho_0^2 + \mu \rho_0,
 \ee
where $x=0,1$. Thus, we can minimize this expression for $x=0,1$ as a function of $\rho_0$, and choose the smallest value. The value of $x$ depends on the sign of $t$ only, whereas the value of $1\ge \rho_0\ge 0$ depends on the values of $U$ and $\mu$. Note that the ground state is very degenerate. Furthermore, note that ${\cal H}^{\rm sym}_{N,1}$ just contains two states (with different parity), namely $|\Omega\rangle$ and $\bar A_1^\dagger|\Omega\rangle$. Any other states, like $|\chi\rangle=\prod_n a^\dagger_{n,1} |\Omega\rangle$, are not invariant under permutations. Therefore, if we minimize the energy density in that space we will not obtain the right result. This explains that if we want to restrict ourselves to symmetric states, we must take density operators,  like for instance, $|\chi\rangle\langle\chi|$, which can indeed be invariant under permutations. As explained above, we can purify those states and indeed work in the space ${\cal H}^{\rm sym}_{N,2}$.

It is illustrative to see how we can obtain the above result with the technique introduced in the previous section. Thus, we have to take a purification corresponding to $M'=2$ modes in each site, with annihilation operators, $a_{n,1}$ and $a_{n,2}$, the original one and the copy. The operators (\ref{barA}) are in this case
 \begin{subequations}
 \bea
 \label{SymModes1}
 \bar A^{(1)}_\mu &=& \sum_{n=1}^N a_{n,\mu}, \\
 \bar A^{(2)} &=& \sum_{n=1}^N a_{n,2}a_{n,1}.
 \eea
 \end{subequations}
The first ones have $k$ odd, whereas the second one has $k$ even. If we restrict ourselves to even parity, we have that the FSBS is composed of linear superpositions of two states,
 \be
 \label{M1FSBS}
 |\psi(c,\alpha)\rangle = |\varphi_0(\alpha)\rangle + c |\varphi_1(\alpha)\rangle,
 \ee
where 
 \begin{subequations}
 \label{varphi01x}
 \bea
 |\varphi_0(\alpha)\rangle &=& e^{\alpha \bar{A}^{(2)\dagger}}|\Omega\rangle,\\
 |\varphi_1(\alpha)\rangle &=& \bar{A}_1^{(1)\dagger} \bar{A}_2^{(1)\dagger} e^{\alpha \bar{A}^{(2)\dagger}}|\Omega\rangle.
 \eea
 \end{subequations}
In fact, in order to simplify the calculations, it is more convenient to take
 \begin{subequations}
 \label{varphi01}
 \bea
 |\varphi_0(\alpha)\rangle &=& \prod_n \left(\cos(\alpha)+\sin(\alpha)a^\dagger_{n,1}a^\dagger_{n,2}\right)|\Omega\rangle,\\
 |\varphi_1(\alpha)\rangle &=& 
 \left(\frac{a^\dagger_{1,1}+a^\dagger_{2,1}}{\sqrt{N}}+ A_1^\dagger\right)
 \left(\frac{a^\dagger_{1,2}+a^\dagger_{2,2}}{\sqrt{N}}+ A_2^\dagger\right)|\varphi_0(\alpha)\rangle,
 \label{varphi01b}
 \eea
 \end{subequations}
where 
 \be
 A_\mu = \frac{1}{\sqrt{N}} \sum_{n=3}^N a_{n,\mu}.
 \ee

In order to determine the energy density, we need to calculate the normalization $\langle\psi(c,\alpha)|\psi(c,\alpha)\rangle$, as well as the expectation value of $\bar H = \hat H /N^2$. Since the state $|\psi\rangle$ is already symmetric with respect to permutations, we just need to take the terms in $\bar H$ that include the first and the second site. Thus we have singled them out in (\ref{varphi01}), since then we will be able to take the limit $N\to\infty$ during the evaluation. Note that then we will be left with a three-site problem, with sites 1, 2, combining the remaining sites into a single one.
Besides, we will have to consider separately the terms with an even number of operators on each site (proportional to $\mu$ and $U$), and those with an odd number (proportional to $t$). In the first ones, we will have to determine the expectation value up to zero order in $1/N$, which will become very simple since we will be able to ignore the terms that go like $1/\sqrt{N}$ in (\ref{varphi01}), and treat the operators $A_\mu$ like simple annihilation operators. In the latter, the zero order in $1/N$ vanishes, in agreement with the discussion of the previous subsections. We will then have to go to order $1/N$, where the first terms in (\ref{varphi01b}) contribute. In order to simplify the notation, we will not write explicitly the dependence on $c$ and $\alpha$, and denote by $\langle X\rangle_0 = \langle \varphi_0|X|\varphi_0\rangle$.

We start determining some expectation values to zero order in $1/N$
 \begin{subequations}
 \bea
 \langle A_\mu A_\mu^\dagger \rangle_0 &=& \cos^2(\alpha),\\
 \langle A_1^\dagger A_2^\dagger \rangle_0 &=& \cos(\alpha)\sin(\alpha),\\
 \langle A_1 A_2^\dagger \rangle_0 &=& 0,\\
 \eea
 \end{subequations}
To the same order, we have
 \be
 \langle A_2 A_1 A_1^\dagger A_2^\dagger \rangle_0 =
 \langle A_1A_1^\dagger \rangle_0 \langle A_2A_2^\dagger \rangle_0 +
 \langle A_2A_1 \rangle_0  \langle A_1^\dagger A_2^\dagger \rangle_0 
 - \langle A_2A_1^\dagger \rangle_0  \langle A_1 A_2^\dagger \rangle_0
  = \cos^2(\alpha),
 \ee
where we have used that the number of terms where four operators act on the same site is of order $N$, so that it vanishes when divided by $N^2$. We see that this is nothing but a Wick's theorem, where we can decompose products of an even number of operators into products of all possible pairs.  

Using the above results, we can easily obtain the normalization
 \be
 \langle\psi|\psi\rangle = 1 + 2 c \sin(\alpha)\cos(\alpha) + c^2 \cos^2(\alpha).
 \ee
Let us now concentrate on the expectation values of the terms of $\bar H$ acting on sites 1 and 2, and that have an even number of operators on each site. According to our prescription, we have to determine them in zeroth order in $1/N$. This means that we can ignore the operators $a_{1,\mu}^\dagger$ and $a_{2,\mu}^\dagger$ in (\ref{varphi01b}), and since the operators $A_\mu$ do not act on sites 1 and 2, we obtain that 
 \begin{subequations}
 \bea
 \langle \varphi_i |a^\dagger_{1,1}a_{1,1}|\varphi_j\rangle &=&
 \langle a^\dagger_{1,1}a_{1,1}\rangle_0 \; \langle \varphi_i |\varphi_j\rangle = \sin^2(\alpha)\langle \varphi_i |\varphi_j\rangle ,\\
 \langle \varphi_i |a^\dagger_{1,1}a_{1,1}a^\dagger_{2,1}a_{2,1}|\varphi_j\rangle &=&
 \langle a^\dagger_{1,1}a_{1,1}a^\dagger_{2,1}a_{2,1}\rangle_0 \; \langle \varphi_i |\varphi_j\rangle = \sin^4(\alpha)\langle \varphi_i |\varphi_j\rangle .
 \eea
 \end{subequations}
The terms of $\bar H$ acting on sites 1 and 2, that have an odd number of operators acting on each site (i.e., proportional to $t$), can be also easily calculated. But now the first non-vanishing order is $1/N$, which will compensate the scaling factor we have added in the Hamiltonian. As before, we have to deal with a three-site problem, and determine the lowest order corrections. We obtain
 \begin{subequations}
 \bea
 \langle \varphi_0 |a^\dagger_{1,1} a_{2,1}|\varphi_0\rangle &=& 0,\\
 \langle \varphi_1 |a^\dagger_{1,1} a_{2,1}|\varphi_1\rangle &=& \frac{\cos^4(\alpha)}{N},\\
 \langle \varphi_0 |a^\dagger_{1,1} a_{2,1}|\varphi_1\rangle &=& \frac{\sin(\alpha)\cos^3(\alpha)}{N},
 \eea
 \end{subequations}

Putting all those results together we get (\ref{E0tU}), with $\rho_0=\sin^2(\alpha)$ and 
 \be
 x=\frac{\left[\sin(\alpha) + c \cos(\alpha)\right]^2}{1+2c \sin(\alpha)\cos(\alpha)+c^2 \cos^2(\alpha)}.
 \ee
Since $\sin^2(\alpha),x\in[0,1]$, we recover the solution given above. 

Before finishing this example, let us notice that the FSBS 
(\ref{M1FSBS}) is, in fact, a Gaussian state. This can be easily shown by noticing that we can always write 
 \be
 |\psi(c,\alpha)\rangle = e^{c \bar{A}_1^{(1)\dagger} \bar{A}_2^{(1)\dagger}} e^{\alpha \bar{A}^{(2)\dagger}} |\Omega\rangle.
 \ee

\subsection{Example: $M=2$}

In the following, we construct for $M=2$ ($M' = 4$) several explicit forms of FSBSs. We will obtain a new class of states, which goes beyond the class of Fermionic Gaussian states. This result motivates a new family of variational states, extending Fermionic Gaussian States, which will be used in the next section to analyze the Hubbard model in 2 spatial dimensions. We will not derive here the ground state energy density - it can be obtained using a tedious, but straightforward procedure, similar to the one of the previous example. 

To simplify notation, we introduce, instead of the operators $\bar A_{\vec \mu}^{(k)}$, the operators $A_{\mu}$ ($k=1$), $B_{\mu_1, \mu_2}$ ($k=2$), $C_{\mu}$ ($k=3$) and $D$ ($k=4$) via
 \begin{subequations}
 \bea
 \label{SymModes}
 A_\mu &=& \sum_{n=1}^N a_{n,\mu}, \quad C_\mu=\sum_{n=1}^N  a_{n,\mu_3}a_{n,\mu_2}a_{n,\mu_1},\\
 B_{\mu_1,\mu_2} &=& \sum_{n=1}^N a_{n,\mu_2}a_{n,\mu_1}, \quad
 D= \sum_{n=1}^N a_{n,4}a_{n,3}a_{n,2}a_{n,1},
 \eea
 \end{subequations}
where $\mu_1<\mu_2<\mu_3$ and $\mu_i\ne \mu$ in the definition of $C$. According to the construction of the previous section (\ref{FSBS}), we can write the FSBSs in terms of those operators. Here we will construct another set of vectors which is more convenient and easy to use, and that will be the basis of the variational states used in the next section.

As shown in (\ref{Symbasis0}) we can build a complete set of states in the symmetric subspace as products of $A^\dagger$'s and $C^\dagger$'s, times powers of $B^\dagger$'s and $D^\dagger$ acting on the vacuum, which we represent as
 \be
 \label{Set1}
 A_i^\dagger \ldots C_j^\dagger \ldots B_k^{\dagger p_k} \ldots D^{\dagger m} |\Omega\rangle.
 \ee
The operators appearing in this expression all commute or anti commute, so that we can choose their ordering arbitrarily. It will be more convenient to work with a different complete set of states of the form
 \be
 \label{Set2}
 A_j \ldots A_i^\dagger \ldots B_k^{\dagger p_k} \ldots D^{\dagger m} |\Omega\rangle.
 \ee
In order to show that Eq. (\ref{Set1}) and Eq. (\ref{Set2}) indeed span the same space, we proceed as follows: We move in (\ref{Set1}) the operator $C^\dagger$ to the left of $D^\dagger$ and use the fact that $C_\mu^\dagger D^{\dagger n}|\Omega\rangle= A_\mu D^{\dagger n+1}|\Omega\rangle /(n+1)$. Then we use the commutation or anti commutation relations between $A_\mu$ and the rest of operators appearing in (\ref{Set1}) to move $A_\mu$ all the way to the left. Those relations read
 \begin{align}
 \{A_{\mu}, A^{\dagger}_k\} &= N \delta_{\mu,k},\\
 \{A_{\mu}, C^{\dagger}_k\}&= \delta_{\mu, k_1}B^{\dagger}_{k_2k_3} - \delta_{\mu, k_2}B^{\dagger}_{k_1k_3} + \delta_{\mu, k_3}B^{\dagger}_{k_1k_2},\\
 \{A_{\mu}, B^{\dagger}_{k_1k_2}\} &=\delta_{\mu k_1}A^{\dagger}_{k_2} - \delta_{\mu k_2}A^{\dagger}_{k_1}.
 \end{align}
Thus, by moving $A$ back to the left we will generate more states, but all of them will be of the same form. We can proceed in the same way with all operators $C^\dagger$. In the end we will arrive at linear combinations of states of the form (\ref{Set2}). As before, instead of using powers of operators $B_j^\dagger$ and $D^\dagger$, we use exponentials so that we can easily take the thermodynamic limit $N\to\infty$, i.e.
 \begin{align}
 \label{eq:prod_to_exp}
 \prod_{\mu_1, \mu_2}(B_{\mu_1,\mu_2}^\dagger)^{p_{\mu_1,\mu_2}}D^{\dagger m} |\Omega\rangle \rightarrow \prod_{\mu_1, \mu_2}e^{\alpha_{\mu_1, \mu_2}B^\dagger_{\mu_1,\mu_2}}e^{\beta D^\dagger}|\Omega\rangle = \prod_n e^{\sum_{\mu_1, \mu_2}\alpha_{\mu_1, \mu_2}a^\dagger_{\mu_1}a^\dagger_{\mu_2}}e^{\beta D^\dagger}|\Omega\rangle.
 \end{align}

We can further replace the exponentials of $B$ by unitary operators, something that will simplify the determination of expectation values, in the following way: Wlog. we can assume that the matrix $(\alpha)_{\mu_1, \mu_2}$ is anti-symmetric and complex. Then, there exists a unitary transformation $w$ such that $(w^T \alpha w)_{2k-1, 2k} = -(w^T \alpha w)_{2k, 2k-1} = \alpha_k$ and $(w^T \alpha w)_{m,n} = 0$ otherwise. Thus, we can define new mode operators $b_{n,\mu_1}^\dagger = \sum_{\mu_2}w_{\mu_1,\mu_2}a^\dagger_{n,\mu_2}$ that fulfill the same anticommutation relations, and rewrite Eq. (\ref{eq:prod_to_exp}) as
 \be
 \prod_{n=1}^N e^{\alpha_1 b_{n,1}^\dagger b_{n,2}^\dagger} e^{\alpha_2 b_{n,3}^\dagger b_{n,4}^\dagger} e^{\beta b_{n,1}^\dagger b_{n,2}^\dagger b_{n,3}^\dagger b_{n,4}^\dagger} |\Omega\rangle.
 \ee
It is immediate to show that one generates the same set of states by replacing
 \be
 e^{\alpha_1 b^\dagger_{n,1} b^\dagger_{n,2}} e^{\alpha_2 b^\dagger_{n,3} b^\dagger_{n,4}}\to e^{\alpha_1' b^\dagger_{n,1} b^\dagger_{n,2}-h.c.} e^{\alpha_2' b^\dagger_{n,3} b^\dagger_{n,4}-h.c.}.
 \ee
where $h.c.$ stands for hermitian conjugate. These latter exponentials are unitary operators which simply implement Bogoliubov transformations (i.e linear canonical transformations). Thus, we can write a set of states which spans the symmetric space as
 \be\label{states_M4}
 |\varphi_{\vec n,\vec m}(u,v,\beta)\rangle= \hat U(u,v) \prod_{i=1}^4 A_i^{n_i} \prod_{j=1}^4
 A_j^{\dagger m_j} e^{\beta D^\dagger} |\Omega\rangle,
 \ee
where $\hat U(u,v)$ is characterized by its action ,
 \be\label{U_M4}
 \hat U(u,v)^\dagger a_{n,\nu} \hat U(u,v) = \sum_{\mu=1}^4 \left[u_{\nu,\mu} a_{n,\nu} + v_{\nu,\mu} a_{n,\nu}^\dagger\right],
 \ee
with $uu^\dagger + v v^\dagger = \Id$, $u v^\dagger + v u^\dagger =0$. Note that we can write
 \be
 \label{states_M4_D}
  e^{\beta D^\dagger}\propto \prod_{n=1}^N \left(1+ \beta \ad_{n,1}\ad_{n,2}\ad_{n,3}\ad_{n,4}\right)|\Omega\rangle,
 \ee
Later,  we will use a class of variational wave functions inspired by (\ref{states_M4_D}) to study the $2D$ Hubbard model with repulsive interaction The FSBS is given by the vectors
 \be
 \sum_{\vec n,\vec m} c_{\vec n,\vec m} |\varphi_{\vec n,\vec m}(u,v,\beta)\rangle,
 \ee
where $c\in \mathbb{C}$ are arbitrary coefficients, and $\vec n,\vec m$ are bit strings.

With the FSBS, we can proceed as in the previous section, namely writing the most general symmetric Hamiltonian of two modes per site, compatible with the parity superselection rule, and then considering only sites 1 and 2. The terms that have an even number of mode operators in sites 1 and 2 give a trivial result, since the normalization factors out and one can use the version of Wick's theorem stated above. For the other terms, one has to consider the three-site problem and expand up to order $1/N$ as before.

\section{Variational wave functions and the Hubbard model}

The simulation of large interacting fermionic many-body systems is one of the big challenges of computational physics. One typically seeks powerful approximation schemes to gain insight into the physical properties of the system. One possibility is the use of appropriate variational wave functions that capture well the physical behavior of the system. A paradigmatic example are fermionic Gaussian states, which can describe a wealth of phenomena, including superconductivity. In particular, they have been successfully applied to simulate the superfluid phase of the attractive Hubbard model in two dimensions within gHFT. In contrast, its application to the repulsive Hubbard model never leads to a paired ground state \cite{Lieb, gHFT}, a feature which is expected to occur in such a model \cite{Micnas}.

In the following we develop a numerical method to determine the ground state energy and correlation functions of an interacting fermionic lattice model in arbitrary dimensions. The method is based on a class of variational wave functions that stem from the FSBS with  $M'=4$. These states can be efficiently described by a number of parameters that scales polynomially in the system size, and allow for an efficient calculation of physical observables. After a general discussion on how to implement a variational algorithm based on these states, we apply them to the two-dimensional spin-full Hubbard model with repulsive interactions. Note that this application is mainly thought as an illustration that aims at demonstrating that these states go well beyond gHFT, since they allow to describe a phase that exhibits at the same time antiferromagnetic correlations and fermionic pairing. Thus, the goal of this Section is not the presentation of en extensive numerical analysis, nor a comparison with other, more developed methods as DMRG or DMFT. The improvement of our methods and its comparison to more sophisticated approaches is the scope of future work.

\subsection{Simulations of Fermionic lattices}

We are interested in understanding the physical properties of fermionic systems on a lattice of $N$ sites, for instance, described by the generic Hamiltonian
\begin{equation}
H = -\sum_k t_{kl} \ad_{k}a_{l} + \sum U_{klmn}\ad_k \ad_l a_m a_n,
\end{equation}
where $t_{kl}, w_{klmn} \in \mathbb{C}$. Such a Hamiltonian describes most of the strongly correlated electronic systems. In particular, it includes the Hubbard model, which is expected to describe the most relevant aspects of high-$T_c$ superconductivity \cite{Hubbard}. We approach the problem in the Majorana picture and define for $k=1, \ldots, 4N$ the operators  $c_{k}= \ad_k+a_k$, $ c_{k+
4N}=(-i)(\ad_k-a_k)$.  These operators obey the canonical anti-commutation relations (CAR) $\{c_k,c_l\}=2 \delta_{kl}$. In this language, the Hamiltonian $H$ is of the form
\begin{equation}\label{eq:Hint}
H = i\sum_{kl} T_{kl}c_k c_l + \sum_{klmn} W_{klmn}c_kc_lc_mc_n ,
\end{equation}
where $T_{kl},W_{klmn} \in \mathbb{R}$. The CAR allow us to antisymmetrize $T$ and $W$ such that $T^T = -T$ while $W$ is antisymmetric under the exchange of any of two adjacent indices.

Our goal is to find a good approximation to the ground state of the Hamiltonian given in Eq.~(\ref{eq:Hint}), as well as its energy. For that, we will take an appropriate family of variational wave functions, inspired by the FSBS for $M' \leq 4$ that we have derived in Sec. IV.C (\ref{states_M4}, \ref{U_M4}, \ref{states_M4_D}). In (\ref{states_M4}), the operators $A_\mu$ and $A_\mu^\dagger$ just involve very few modes (4 out of $N$), so that for large systems they will not give noticeable consequences, which drives us to take $\vec n=\vec m=0$. Now, for $\beta =0$, the FSBS include states that are obtained from that vacuum by Bogoliubov transformations generated by the unitary $\hat U(u,v)$. If we relax now the constraint that $\hat U$ is of the form (\ref{U_M4}), but instead allow for an arbitrary Bogoliobov transformation involving all sites, we arrive at the class of fermionic Gaussian states, which has proven successful in different context. Thus, it is natural to do the same procedure with the term (\ref{states_M4_D}) and allow for arbitrary exponentials of products of any sets of four creation operators (in whatever basis). It is relatively simple to convince oneself that the set of states one obtains at the end is
 \begin{subequations}
 \begin{align}
 \label{eq:variationalstates}
 |\phi_{\vec \beta, \gamma}\rangle &= U_\gamma|\psi_{\vec \beta}\rangle,\\
 |\psi_{\vec \beta}\rangle &= \prod_{n=1}^m \left(\cos \beta_n + \sin \beta _n \ad_{n,1}\ad_{n,2}\ad_{n,3}\ad_{n,4}\right)|\Omega\rangle \label{eq:vec_psi}
 \end{align}
 \end{subequations}
These states are normalized and parametrized by $8N^2 - N$ parameters. Some of them ($4N(4N-1)/2$) parametrize the Bogoliubov transformation $\hat U_{\gamma} = e^{iH_\gamma}$, where $H_\gamma = H_\gamma^\dagger = i\sum_{kl}(\gamma)_{kl}c_kc_l$ and are collected in the real and anti symmetric matrix $\gamma$. The rest ($N$) are collected in the vector $\vec \beta$ and describe the state (\ref{eq:vec_psi}). Note that, despite the fact that our variational ansatz is inspired by the permutationally and rotationally invariant states, ultimately we relax these symmetries, and in particular break the translational symmetry, by considering states with local Bogoliubov transformation and with local terms of the form (\ref{eq:vec_psi}). It is precisely the breakdown of the translation symmetry that allows us to describe in a robust way antiferromagnetic order, as discussed below.

Before we explain how to apply these wave functions variationally, let us discuss a simple interacting model for which the states defined in Eq. (\ref{eq:variationalstates}) are the exact ground state. Consider the Hamiltonian
\begin{align}
H_{4} =  -\sum_{k,l}t_{k,l, \sigma, \sigma'}\ad_{k, \sigma} a_{l\sigma'} + U \sum_{n} \ad_{n,1}\ad_{n,2}\ad_{n,3}\ad_{n,4}+ h.c.
\end{align}
For $U=0$, the ground state is a Gaussian state, i.e. a state where $\beta_n = 0$ for all $n$. For $t=0$, on the other hand, the ground state is 
\begin{align}
|\Psi\rangle = \bigotimes_n = \frac{1}{\sqrt{2}}\left(\ad_{n,1}\ad_{n,2} \pm \ad_{n,3}\ad_{n,4} \right)|\Omega \rangle,
\end{align} 
where $\pm$ depends on the sign of $U$. It is straightforward to show that this state is of the form (\ref{eq:variationalstates})  with $\beta_n = \pm \pi/4$ and $U_\gamma = e^{\pi/2\sum_n \ad_{n,1}\ad_{n,2}  - a_{n,2}a_{n,1} }$. Thus, our new ansatz states will describe well the physics of such a system, at least in the regime of weak and strong interaction.

In the following we develop a numerical technique that allows us to find the ground state within the variational class of states defined in Eq. (\ref{eq:variationalstates}). The state with minimal energy can be obtained by solving the minimization problem
\be\label{eq:minimizeE}
E_{\min} =\min_{\vec \beta, \gamma} E(\vec \beta, \gamma) =  \min_{|\phi_{\vec \beta, \gamma}\rangle} \langle \phi_{\vec \beta, \gamma}|H   |\phi_{\vec \beta, \gamma}\rangle.
\ee

We define the real one-and two-body correlation matrices
 \begin{subequations}
 \begin{align}
 G_{kl}^{(\vec \beta, \gamma)} &= \frac{i}{2}\langle \phi_{\vec \beta, \gamma}| [c_k,c_l]|\phi_{\vec \beta, \gamma}\rangle,\\
 K_{klmn}^{(\vec \beta, \gamma)} &= \frac{1}{4!}\langle \phi_{\vec \beta, \gamma}| [[c_kc_lc_mc_n]]|\phi_{\vec \beta, \gamma}\rangle.
 \end{align}
 \end{subequations}
Here, $[,]$ denotes the usual commutator, and the symbol $[[...]]$ denotes the complete antisymmetrization of the operator $c_kc_lc_mc_n$. These matrices uniquely define the state, and depend on a number of parameters that scales only polynomially in the system size. Furthermore, $G$ and $K$ have a very efficient description in the following way: The unitary matrix $U_{\gamma}= e^{iH_\gamma}$  is a  Bogoliubov transformation on the mode operators, which realizes an orthogonal transformation on the Majorana operators via $\hat U^\dagger_\gamma c_k \hat U_\gamma = \sum_l (O_\gamma)_{kl}c_l$, where $O_\gamma = e^{\gamma}$. Then, we introduce the matrices
\begin{align}
G_{kl}^{(\vec \beta, 0)} = \frac{i}{2}\langle \psi_{\vec \beta}| [c_k,c_l] |\psi_{\vec \beta}\rangle,\;\;\;\; K_{klmn}^{(\vec \beta, 0)} = \frac{1}{4!}\langle \psi_{\vec \beta}| [[c_kc_lc_mc_n]]|\psi_{\vec \beta}\rangle,
\end{align}
that are related to $G^{(\vec \beta, \gamma)}$ and $K^{(\vec \beta, \gamma)}$ via
\begin{align}
G^{(\vec \beta, \gamma)} = O_\gamma G^{(\vec \beta, 0)}O_\gamma^T,\;\;\;\;K^{(\vec \beta, \gamma)}  &= (O_\gamma \otimes O_\gamma)K^{(\vec \beta, 0)}(O_\gamma^T \otimes O_\gamma^T).
\end{align}

Furthermore, using Eq. (\ref{eq:variationalstates}), it is straightforward to see that the matrices $G^{(\vec \beta, 0)}$ and $K^{(\vec \beta, 0)}$ are sparse and depend only on the $N$ parameters $\beta_n$, and can thus be represented efficiently (see Appendix C). Then, using Eqs. (\ref{eq:Hint}) we can formulate the minimization of the energy as an optimization problem of the form
 \begin{align}\label{eq:variational_energy}
E_{\min}(\vec \beta, \gamma) &= \min_{\vec \beta, \gamma}\left(-\mathrm{tr}[G^{(\vec \beta, \gamma)}T]+ \mathrm{tr}[K^{(\vec \beta, \gamma)} W]\right)\nonumber\\
&=\min_{\vec \beta, \gamma}\left( -\mathrm{\tr}[G^{(\vec \beta, 0)}O_{\gamma}^TTO_{\gamma}]+ \mathrm{\tr}[K^{(\vec \beta, 0)} (O_{\gamma}^T\otimes O_{\gamma}^T)W(O_{\gamma}\otimes O_{\gamma})]\right).
\end{align}
Minimizing Eq. (\ref{eq:variational_energy}) requires the optimization over $4N(4N-1)/2$ parameters for $O_\gamma$ and $N$ parameters for $\vec \beta$. For large systems, this problem can be attacked as follows: Introduce a time dependence in the parameters  $\vec \beta (t)$, $\gamma (t)$. Then, starting from an arbitrary initial configuration  $\vec \beta (t), \gamma (t)$ with energy $E(\vec \beta (t), \gamma (t))$ and given a time interval $\delta t$, we want to find $\vec \beta (t+\delta t), \gamma (t+\delta)$ such that $E(\vec \beta (t+\delta t), \gamma (t+\delta t)) <  E(\vec \beta (t), \gamma (t)) $. As it is shown in Appendix C, this can be achieved in the following way: First, keep $\vec \beta(t)$ fixed and expand $\gamma(t+\delta t) = \gamma(t) + h_{\gamma} \delta t$, where
\begin{equation}\label{eq:A}
h_\gamma = [T,G^{(\vec \beta, \gamma)}] - 2\mathrm{tr}_2 [[W,K^{(\vec \beta, \gamma)}]]= - h_\gamma^T.
\end{equation}
Here, $\mathrm{tr}_2 [W K]_{kl} = W_{kxyz}K_{xyzl}$.  As we show in Appendix C, the energy decreases under this operation. Next, we keep $\gamma(t+\delta t)$ fixed, and minimize over $\vec \beta (t)$. This can be implemented numerically, e.g., via a gradient optimization. The energy is always decreasing under these operations, and for $t \rightarrow \infty$ we arrive at the ground state which is completely described in terms of the matrices $G_{\infty} = G^{\left(\beta(t \rightarrow \infty), \gamma(t \rightarrow \infty)\right)}$, $K_{\infty} = K^{\left(\beta(t \rightarrow \infty), \gamma(t \rightarrow \infty)\right)}$. Having these matrices at hand, we also have all one-and two-body correlation functions available, and higher order correlation functions can be computed in a straightforward way.

\subsection{Application: The two-dimensional Hubbard model}
In the following we apply the numerical method that we have derived in the last Section to the two-dimensional Hubbard model on a square lattice,
\begin{equation}
H= -t\sum_{ \langle x,y  \rangle \in \Lambda, \sigma}\ad_{x,\sigma}a_{y,\sigma}+ U \sum_{x\in \Lambda} \left(n_{x\uparrow}-\frac{1}{2}\right)\left(n_{x\downarrow}-\frac{1}{2}\right) + \mu
\sum_{x,\sigma}n_{x,\sigma},\label{eq:Hubbard_H_2d}
\end{equation}
where $x$ and $y$ are points on a
two-dimensional lattice, $\langle x,y \rangle$ denote
nearest-neighbors and $\sigma = \uparrow, \downarrow$ denotes the spin degree of freedom. We consider only the case where the number of spin-up and spin-down particles are equal, so that we can use the same chemical potential for the two species. For $U < 0$ $(U > 0 )$ the second term in $H$ is an attractive (repulsive) on-site interaction between particles of
opposite spin. In the following, we set $t=1$ as the energy scale of the system. Note that the case of half-filling is characterized by $\mu = 0$.

Despite its simple structure the Hubbard model realizes a wide class of non-trivial phases. In the case of an attractive interaction the system is in a superfluid state that can be described well by the BCS wave functions \cite{Micnas, Scalapino} included in the framework of gHFT.  For a repulsive interaction at half filling, the system is predicted to be in a spin ordered phase. This is numerically confirmed by Quantum Monte Carlo simulations \cite{QMC1, QMC2}, and we have shown in \cite{gHFT} that gHFT can be used to obtain the same results. Less is known away from half-filling, where QMC can no longer be applied due to the notorious sign problem. In this regime, a paired superfluid phase with $d$-wave symmetry is predicted (see e.g. \cite{highTc1, highTc2}). Thus the Hubbard model with repulsive interactions has undergone a wide investigation during the last years (see \cite{HubbNum1, HubbNum2,HubbNum3,HubbNum4,HubbNum5,HubbNum6,HubbNum7,HubbNum8,HubbNum9,HubbNum10,HubbNum11} and references therein). Since the gHFT solution for $U>0$ is always unpaired \cite{Lieb}, this approach does not allow to verify this prediction.

In the following we use the variational states defined in Eq. (\ref{eq:variationalstates}) to get insight into the ground state properties of the two-dimensional repulsive Hubbard model with doping on a $10 \times 10$ square lattice with periodic boundary conditions. As an example, we consider three sets of parameters, $(U, \mu) = (3, -0.5), (5, -1), (6, -1.5)$ and arrive at the following results: (i) In all three cases the minimal energies $E_{FSBS_4}$ are below the gHFT energies $E_{\mathrm{gHFT}}$. (ii) For the case $(U, \mu) = (5,-1)$ and $(U, \mu) = (6, -1.5)$ the system is in a state with anti-ferromagnetic order. (iii) The system is paired for all three sets of parameters. Let us discuss these results in more detail.
\begin{table}[t]
\begin{center}
\begin{tabular}{| c| c | c | c |}
\hline
$(U, \mu)$ &   $(3, -0.5)$    & $(5, -1)$      & $(6, -1.5)$\\\hline
$N_{\mathrm{tot}}$                 &  $74$ & $80$  & $98$\\\hline
$E_{\mathrm{gHFT}}$ &  $-152.4284$ & $-166.9931$ & $-206.804$ \\\hline
$E_{FSBS_4}$ & $-155.0148$ & $-172.1345$ & $-209.3691$ \\\hline
$\mathcal{M}(\rho)$ & 1.2 & 1.3& 1.1\\\hline
\end{tabular}
\end{center}
\caption{Numerical results for the $2d$ Hubbard model with repulsive interaction $U$ and chemical potential $\mu$ on a $10 \times 10$ lattice. We present the total particle number $N_{\mathrm{tot}}$, the minimal variational energy obtained in gHFT, $E_{\mathrm{gHFT}}$, the minimal variational energy obtained for the variational states obtained by using the FSBS for $M'=4$ defined in Eq. (\ref{eq:variationalstates}), $E_{FSBS_4}$ as well as the pairing $\mathcal{M}(\rho)$ (cf. Eq. (\ref{eq:pairing})) obtained using the FSBS for $M'=4$. The results clearly show that the use of the variational states (\ref{eq:variationalstates}) leads to an decrease in the variational energy compared to gHFT. Further, a value of the pairing measure $\mathcal{M}(\rho)>1$ indicates that the FSBS for $M'=4$ allow to describe a paired phase in the repulsive regime of the Hubbard model, a feature that is absent in gHFT.}
\end{table}
First, we calculated the ground states within the family of variational states (\ref{eq:variationalstates}). After a fast decrease of the energy at the beginning, the algorithm shows slow convergence. In Table I we present results for $\mathcal{O}(5\cdot10^4)$ iterations. The results clearly show that the new ansatz allows us to achieve smaller energies than using gHFT.At this point we would also like to comment on the performance of our algorithm. At the current stage, the algorithm shows slow convergence, and we are currently working on improving the method to make it more appropriate for an efficient application to large system.

\begin{figure}[t]
\begin{center}
\includegraphics[width = 0.85 \columnwidth]{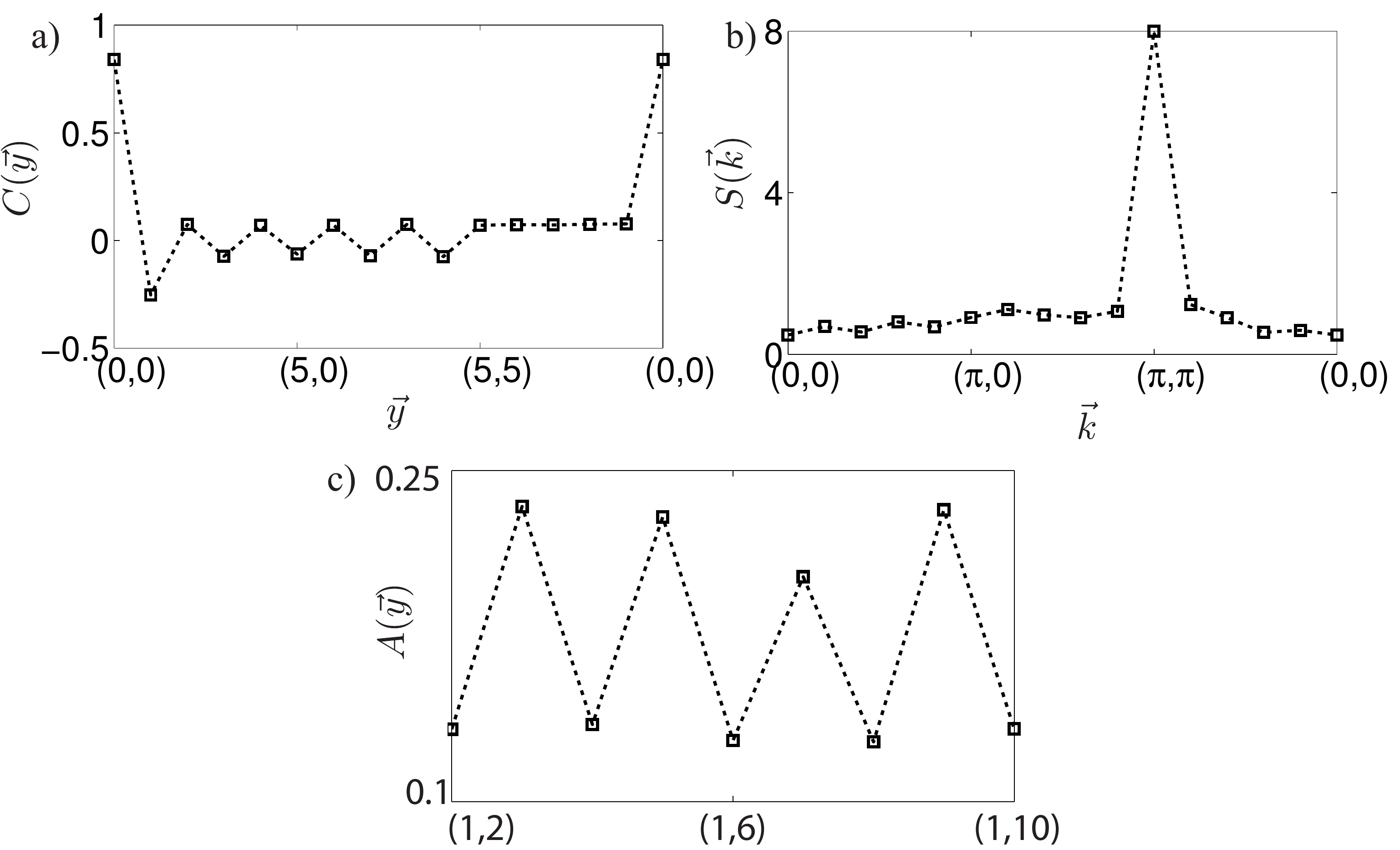}
\end{center}
\caption{Numerical results of various correlation functions for $(U, \mu) = (6, -1.5)$. a) Equal-time spin-spin correlation function $C(\vec y) = \langle (n_{(\vec x + \vec y )\uparrow} - n_{(\vec x + \vec y )\downarrow})(n_{\vec x\uparrow} - n_{\vec x \downarrow})\rangle$. b) Magnetic structure factor $S(\vec k) = \sum_{\vec x}e^{i\vec k \cdot \vec x}C(\vec x)$. The sharp peak at $(\pi, \pi)$ is an indicator for the anti-ferromagnetic order inherent in the system. c) Order parameter for anti-ferromagnetic order, $A(\vec y ) = \langle n_{\vec x\uparrow}n_{(\vec x + \vec y)\downarrow}\rangle$.\label{fig_positive_u}}
\end{figure}

Second, we study the magnetic properties of the system by investigating the equal-time spin-spin correlation function $C(\vec y) = \langle (n_{(\vec x + \vec y )\uparrow} - n_{(\vec x + \vec y )\downarrow})(n_{\vec x\uparrow} - n_{\vec x \downarrow})\rangle$, the magnetic structure factor $S(\vec k) = \sum_{\vec x}e^{i\vec k \cdot \vec x}C(\vec x)$ and the order parameter for anti-ferromagentic order, $A(\vec y ) = \langle n_{\vec x\uparrow}n_{(\vec x + \vec y)\downarrow}\rangle$.  As an example, we show results for $(U, \mu) = (6, -1.5)$  in Fig. \ref{fig_positive_u}. Similar results are obtained for $(U, \mu) = (5,-1)$ and clearly indicate the presence of anti-ferromagnetic order in the system.

Third, we address the question if our ground states exhibit pairing. The appearance of pairing would support the hypothesis that the repulsive Hubbard model can support superconductivity. We use the pairing measure defined in Ref.~\cite{Pairing} to give an answer to this question. There, it has been shown that for any unpaired state of $N_{tot}=2N$ particles and $4 M$ modes
\be\label{eq:pairing}
\mathcal{M}(\rho)  = \frac{1}{N}\max_{\{\ad_i\}_i}\sum_{k,l=1}^{2M}|\langle \ad_{2k-1} \ad_{2k}a_{2l}a_{2l-1}\rangle_{\rho}| \leq 1,
\ee
where we optimize over all possible bases of modes $\{\ad_i\}_i$ . For all three sets of parameters we find that the ground states are paired, as shown in Table I (see Appendix D for the numerical implementation of the minimization).

Summarizing, the variational wave functions obtained form the FSBS with $M'=4$ lead to smaller energies compared to gHFT. Further, the variational ground states describe paired phases with anti-ferromagnetic order for the repulsive Hubbard model, a feature that is elusive within gHFT.

\section{Conclusion}
In this work we have investigated the ground states of fermionic lattice systems of $N$ sites and $M$ modes per site with permutation symmetry in the limit $N \rightarrow \infty$. We have explained that, opposed to the case of spin systems, a careful scaling analysis of the typical terms appearing in the Hamiltonian is necessary in order to obtain a result that captures the rich physics of fermionic many-body systems. Then we have introduced the \emph{Fermionic Symmetric Basic States (FSBS)} from which the ground states of such Hamiltonians can be easily computed. Such a result can be seen as a quantum de Finetti theorem for fermionic lattice systems. In contrast to its spin version, where such a set is simply the set of product states, the fermionic states obtained in this way are highly non-trivial and their form depends on the number $M$ of modes per site. In the case of $M=2$, the FSBS are contained within the set of fermionic Gaussian states that have proven as powerful variational wave function within generalized Hartree Fock Theory. For $M>2$ however, the FSBS are no longer contained in the set of fermioninc Gaussian states, but go beyond. As an example, we have considered the case of $M=4$ modes per site and constructed the FSBS explicitly. Having this result at hand, we have applied those states as a variational class of states to construct a numerical technique that allows for an efficient simulation of the ground state and the ground state properties of interacting fermionic lattice systems in arbitrary dimensions and geometries. The algorithm depends on a number of parameters that scales only polynomially in the system size and allows to calculate of correlation functions efficiently. To test our technique, we have applied it to the repulsive Hubbard model on a two-dimensional $10\times 10$ square lattice with doping, a scenario where gHFT fails to capture the predicted superconducting phase, and QMC cannot be applied due to the fermionic sign problem. We find that the new class of variational wave functions leads to lower energies than gHFT, and supports paired phases with anti-ferromagnetic correlation functions.

Note, that the purpose of this numerical analysis performed in this work is to show that the FSBS lead to a class of variational trial states that goes beyond the standard approach of gHFT and allows to capture physical properties of the system that are elusive in gHFT. In this respect, our results should be seen from a qualitative point of view, while a systematic comparison with other numerical techniques is out of the scope of the present work. Another direction is to systematically investigate the FSBS and the related variational wave functions for $M>4$, to benchmark them with physically relevant models and to compare the with other numerical approximation schemes. This might lead to new numerical techniques that allow to gain insight into the physics of interacting fermionic lattice systems  that are hard to capture otherwise.

\section{Acknowledgements}
This work has been supported by the EU project SIQS, the ERC AdG QUAGATUA, and the MINCIN Project TOQATA. ML thanks the Alexander von Humboldt Foundation for support. JIC thanks F. Brandao for useful discussions.

\section*{Appendix A: Second quantization formalism for spin problems}

As we mentioned in the introduction, we can also use second quantization to solve the problem posed in Section II. Denoting as before by $\{|n\rangle\}_{n=1}^{K'}$ a single-spin basis, we construct the Fock space as usual, by defining $a_n^\dagger$ as the bosonic operator that creates a spin in the state $|n\rangle$ out of the vacuum $\Omega$. For instance,
 \be
 |\phi\rangle^{\otimes N}\to a(\phi)^{\dagger N}|\Omega\rangle,
 \ee
where
 \be
 a(\phi)=\sum_{n=1}^{K'} c_n a_n,
 \ee
and we have ignored the normalization constant. In this language, an arbitrary symmetric Hamiltonian with up to two-spin interactions, where the spins are $K$-level systems, can be written as
 \be
 H = \sum_{n_i=1}^K h_{n_1,n_2,n_3,n_4} a^\dagger_{n_1}a^\dagger_{n_2}a_{n_3}a_{n_4}.
 \ee
The tensor $h_{n_1,n_2,n_3,n_4}$ is invariant under the exchange of indices $n_1\leftrightarrow n_2$, $n_3\leftrightarrow n_4$, and hermitian in the joint-indices $n_{1,2}$ and $n_{3,4}$. In principle, we could include a term quadratic in creation and annihilation operators as well, which would account for single-spin terms in the original Hamiltonian. However, since
$[\hat N,H]=0$ with
 \be
 \hat N=\sum_{n=1}^K a^\dagger_n a_n,
 \ee
the total number of spins, such term can be included in the tensor $h$. We are interested in determining the minimum energy for all possible tensors with the mentioned properties, and for $\langle \hat N \rangle=N\to \infty$.

The above problem is very-well known in the context of many-body physics, and it is very much related to the phenomenon of Bose-Einstein condensation \cite{Stringari}. In such case, one usually adds a "chemical potential" to the Hamiltonian, namely a term $\mu \hat N$, and minimizes the energy without any restriction. In fact, the value of $\mu$ fixes a particular value of $N$, so that the desired limit can be obtained by taking the appropriate limiting procedure in $\mu$. In order to find the lowest energy, one chooses the set of coherent states
 \be
 \label{coherent}
 |\Psi(\vec\alpha)\rangle = e^{\sum_{n=1}^{K'} \alpha_n a_n^\dagger} |\Omega\rangle,
 \ee
and applies the variational method (note that we should take a purification of the state by doubling the number of modes, but we will not do that here to simplify the presentation). Note that the state $|\phi\rangle^{\otimes N}$ and the state (\ref{coherent}) with $\alpha_n=\sqrt{N} c_n$ give the same expectation values in the limit $N\to \infty$, and thus one obtains the right result in this way. Furthermore, one can obtain corrections to the de Finetti result \cite{Christandl}, by following the procedure used in the study of Bose-Einstein condensation. That is, once one has obtained $c_n$, one can displace the Hamiltonian $H\to \hat D(\alpha) H \hat D(\alpha)^\dagger$, and expand the result up to second order in the creation and annihilation operators, from which one can determine the ground state energy. Here $\hat D$ is the displacement operator that transforms $|\Psi(\vec\alpha)\rangle\to |\Omega\rangle$. Note that this procedure gives corrections in $1/N$. Alternatively, noticing that the state obtained from this procedure is Gaussian (since it is the displaced state of the ground state of a quadratic Hamiltonian), one can obtain a more accurate description by minimizing the energy of $H$ with respect to the set of Gaussian states. This establishes the connection of the de Finetti theorem and Gaussian states mentioned in the introduction.

\section*{Appendix B: Lattices in infinite dimensions}

In this appendix we relate the problem studied in Sections II and III to the problem of lattices in infinite dimensions. As we will show, for spin systems the solution of both coincides, whereas for Fermions, the first provides upper and lower bounds to the second problem.

We consider a Hamiltonian $H_d$, in $d$ spatial dimensions, with $N=(2\ell+1)^d$ lattice sites. Each site is characterized by a lattice vector, $\vec i$ which is a $d$-component vector with each of the components running from $-\ell,-\ell+1,\ldots \ell$. We will assume two-site interactions as before, so that
 \be
 H_d = \sum_{\langle \vec i,\vec j\rangle} h_{\vec i,\vec j},
 \ee
where $\langle \vec i,\vec j \rangle$ denotes nearest-neighbors ($|\vec i-\vec j|=1$), and they contain the appropriate factors of $N$, as discussed in Sections II and III such that they give a non-trivial contribution in the limit procedures we will consider. We will also assume translation and rotation symmetries. For this purpose, we define the translation and rotation operators, $T_{\vec k}$ and $R_{x,y}$, respectively, and assume periodic boundary conditions. They fulfill $T_{\vec k} X^\alpha_{\vec i} T_{\vec k}^\dagger= X^\alpha_{\vec i+\vec k}$ and $R_{x,y} X^\alpha_{\vec i} R_{x,y}^\dagger=X^\alpha_{\vec i'}$, where $\vec i,\vec k$ are lattice vectors, $x,y=1,\ldots,d$, $x\ne y$, $i'_x=i_y$, $i'_y=-i_x$ and $i'_z=i_z$ for all $z\ne x,y$. Thus, $H_d=T_{\vec k} H_d T_{\vec k}^\dagger= R_{x,y} H_d R_{x,y}^\dagger$. We will denote by ${\cal D}^{\rm sym}_{N,K}$, the set of density operators that are translational and rotational invariant, i.e.,
 \be
 \label{RhoSym}
 \rho=T_{\vec k} \rho T_{\vec k}^\dagger= R_{x,y} \rho R_{x,y}^\dagger.
 \ee

We will be interested in the ground state energy density
 \be
 \label{EMin}
 E_d= \lim_{d\to\infty} \frac{1}{(2\ell +1)^d} \min_{||\rho||_1} {\rm tr}(\rho H_d),
 \ee
in the limit $d\to \infty$ (keeping $\ell<\infty$). Now we will derive an upper and a lower bound to $E_d$ in terms of states invariant under permutations, which can be obtained with the methods of Sections II and III.

 \begin{figure}[t]
 \begin{center}
 \includegraphics[width = 0.7 \columnwidth]{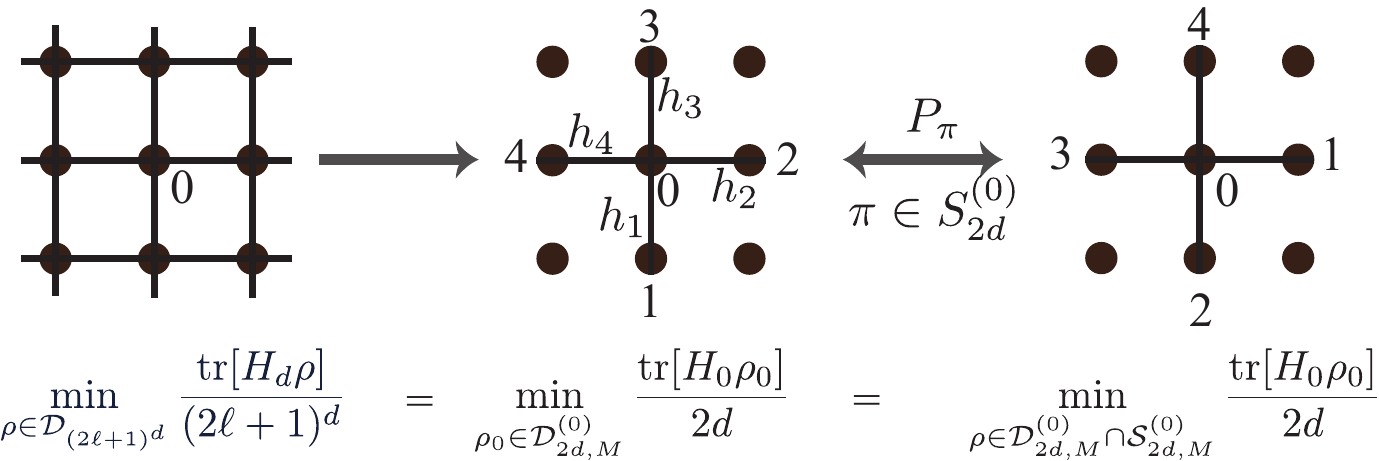}
 \end{center}
 \caption{Sketch illustrating the derivation behind the calculation of the minimal mean energy for $H_d$.\label{fig:SymmetricHamiltonians}}
 \end{figure}

The upper bound is trivial, as we can just restricting the minimization to the set of SSBS or FSBS,
 \be
 \label{EMin}
 E_d\le  E^{\rm up}=\lim_{d\to\infty} \frac{1}{(2\ell +1)^d} \min_{|\Psi\rangle \in {\rm S(F)SBS}} \langle\Psi|H_d|\Psi\rangle.
 \ee
For the lower bound, we proceed as follows (see Fig. \ref{fig:SymmetricHamiltonians}). Due to the symmetry assumption, we can take $\rho\in {\cal D}^{\rm sym}_{N,K}$ in the minimization of Eq. (\ref{EMin}). Therefore, we can also write
 \be
 \label{minim}
 E_d=\frac{1}{2d} \min_{\rho_0\in {\cal D}^{(0)}_{2d,K}} {\rm tr}(H_0 \rho_0),
 \ee
where $H_0$ contains only those terms of $H_d$ which include the site ${\vec 0}$, that from now will be referred to as site 0. Thus,
 \be
 H_0 = \sum_{n=1}^{2d} h_n,
 \ee
where $n$ now enumerates the nearest neighbors of site 0, and $h_n$ is the operator in $H_d$ that is acting on the sites 0 and $n$. We have denoted by ${\cal D}^{(0)}_{2d,K}$ the set of density operators that are acting on site $0$ and all its neighboring sites, and which can be obtained as reduced states out of any $\rho\in {\cal D}^{\rm sym}_{(2\ell+1)^d,K}$. The Hamiltonian $H_0$ is obviously invariant under permutations of any pairs of neighboring sites of 0. Thus, we can restrict the minimization in (\ref{minim}) to $\rho_0 \in {\cal D}^{(0)}_{2d,K} \bigcap{\cal S}_{2d,K}^{(0)}$, where the later is the convex set of density operators (also acting on the site $0$ and all its neighbors) that are invariant under any permutation of the sites $1,\ldots,2d$ (but not 0). By relaxing this condition on $\rho_0$, we obtain the lower bound
 \be
 \label{Elow}
 E_\infty \ge E^{\rm low} = \min_{\sigma\in {\cal S}_{K}^{(0,1)}} {\rm tr}(\sigma h_1).
 \ee
Here,
 \be
 {\cal S}_{K}^{(0,1)} = \lim_{d\to \infty} {\cal S}_{2d,K}^{(0,1)},
 \ee
and ${\cal S}_{2d,K}^{(0,1)}$ is the set of density operators acting on sites 0 and 1, and which are reduced from a density operator in ${\cal S}_{2d,K}^{(0)}$.

For spin lattices, one can easily show that the two bounds coincide, and thus product states solve the lattice problem in infinite dimensions as well. This can be shown by using a similar procedure to the one utilized in Section II. We define
 \be
 |\Psi\rangle = \int d\mu_{\phi_0} d \mu_{\phi} f(\phi_0, \phi)|\phi_0\rangle|\phi\rangle^{\otimes 2d} ,
 \ee
where $\phi_0\in {\cal H}_{K}$ and $\phi\in {\cal H}_{K'}^{\otimes 2}$, which is symmetric with respect to permutations of all spins except the first one, $|\phi_0\rangle$. If we calculate the reduced density operator of particle 0 and 1, and take the limit $N\to\infty$ we obtain
 \be
 \sigma \to \int d\mu_{\phi_0}d\mu_{\phi_0'}d\mu_\phi \bar f(\phi_0',\phi)f(\phi_0,\phi) |\phi_0\rangle\langle\phi_0'|\otimes|\phi\rangle\langle\phi|,
 \ee
which is again separable, although not necessarily symmetric with respect to the exchange of spins 1 and 2. We can again use it to determine the lower bound $E^{\rm low} $ in (\ref{Elow}). Since $h_1$ is symmetric with respect to particles 0 and 1, we obtain that $E^{\rm low}=E^{\rm up}=E_\infty$.

\section*{Appendix C: Derivation of Eq. (\ref{eq:A})}

In this appendix we explain how to arrive at Eq. (\ref{eq:A}) of the main text. There,  it has been shown that the minimization problem can be reformulated as
\be
 E(\vec \beta, \gamma) = \min_{\vec \beta, \gamma} -\mathrm{tr}[G^{(\vec \beta, 0)}O_{\gamma}^TTO_{\gamma}]+ \mathrm{tr}[K^{(\vec \beta, 0)} (O_{\gamma}^T\otimes O_{\gamma}^T)W(O_{\gamma}\otimes O_{\gamma})].
\ee
Recall that
\begin{align}
G_{kl}^{(\vec \beta, 0)} &= \frac{i}{2}\langle \psi_{\vec \beta}| [c_k,c_l] |\psi_{\vec \beta}\rangle,\\
K_{klmn}^{(\vec \beta, 0)} &= \frac{1}{4!}\langle \psi_{\vec \beta}| [[c_kc_lc_mc_n]]|\psi_{\vec \beta}\rangle,
\end{align}
$O_{\gamma}$ is an orthogonal transformation, and $|\psi_{\vec \beta}\rangle = \prod_{n=1}^m \left(x_n + y_n \ad_{n,1}\ad_{n,2}\ad_{n,3}\ad_{n,4}\right)|\Omega\rangle$, where $x_n = \cos \beta_n$, $y_n = \sin \beta_n$. We show first that the matrices $G_{kl}^{(\vec \beta, 0)}$ and $K_{klmn}^{(\vec \beta, 0)}$ allow for an efficient representation in terms of $x_n$ and $y_n$. It is easy to show that
\begin{equation}
G^{(\vec \beta, 0)} = \bigoplus_{n = 1}^N (1- 2|y_n|^2) \left(\begin{array}{cc} 0 & \mathbb{I} \\
-\mathbb{I} & 0      \end{array}\right).
\end{equation}
Further, the tensor $K^{(\vec \beta, 0)}$ can be calculated in the following way: Consider first the case where $i,j,k,l \in \{(n,1), (n,2), (n,3), (n,4)\}$. Then $K^{(0)} = K_1 + x_ny_n K_2 $, where $K_1$, $K_2$ are constant tensors that are the same for each $n$. Now, if $i,j,k,l$ do not all belong to the set $\{(n,1), (n,2), (n,3), (n,4)\}$, we can apply Wick's theorem and obtain $K^{(0)}_ {ijkl} = -G^{(\vec \beta, 0)}_{ij}G^{(\vec \beta, 0)}_{kl} + G^{(\vec \beta, 0)}_{ik}G^{(\vec \beta, 0)}_{jl} - G^{(\vec \beta, 0)}_{il}G^{(\vec \beta, 0)}_{jk}$. Thus, $G$ and $K$ are sparse.

To perform the minimization of the energy, we perform first, for fixed $\vec \beta$, the optimization over the parameters $\gamma$ that describe the orthogonal matrix $O_\gamma$. We define a Lagrangian
\begin{eqnarray}
L(\vec \beta, \gamma) =E(\vec \beta, \gamma) - \sum_{i,j}\lambda_{i,j}(O_\gamma O_\gamma^T -\mathbb{I})_{i,j},
\end{eqnarray}
with Lagrangian multipliers $\lambda_{ij}$. We take derivatives with respect to $O_\gamma$, use $\mu =  \mu^T$ and recall the definitions $G^{(\vec \beta, \gamma)} = O_{\gamma}G^{(\vec \beta, 0)}O_\gamma^T$, $K^{(\vec \beta, \gamma)}  = (O_\gamma \otimes O_\gamma)K^{(\vec \beta, 0)}(O_\gamma^T \otimes O_\gamma^T)$. This leads to the following necessary condition for a minimum:
\begin{equation}\label{eq:cond_energy}
[G^{(\vec \beta, \gamma)},T]- 2\mathrm{tr}_2 [[W^{(\vec \beta, \gamma)},K]] = 0,
\end{equation}
where $\mathrm{tr}_2 [WK^{(\vec \beta, \gamma)}]_{vz} = W_{vjkl}K^{(\vec \beta, \gamma)}_{klzj}$. This matrix equation is hard to solve for large system sizes. However, as we show now, we can solve this equation in the following way: We linearize $\gamma(t + \delta t) = \gamma(t) + \delta t h_{\gamma}$, so that  $O_{\gamma(t+\delta t)} = e^{h_{\gamma} \delta t} O_{\gamma(t)}$. For small $\delta t$, we can write $O_{\gamma(t+\delta t)}= (\mathbb{I} + h_{\gamma} \delta t) O_{\gamma(t)} + \mathcal{O}(\delta t^2)$. Then $\delta E(t) = E(t+\delta t) - E(t)$ is given by
\begin{eqnarray}\label{eq:deltaE}
\delta E(t) &=& - (\mathrm{tr}[[h_{\gamma},G^{(\vec \beta, \gamma)}]T] - \mathrm{tr}[(\mathbb{I}\otimes h_{\gamma} + h_{\gamma} \otimes \mathbb{I})K^{(\vec \beta, \gamma)} (\mathbb{I}\otimes h_{\gamma} + h_{\gamma} \otimes \mathbb{I})U])\delta t + \mathcal{O}(\delta t^2)\nonumber\\
&=&2\mathrm{tr} \left[h_\gamma\left([T,G^{(\vec \beta, \gamma)}] - 2\tr_2 [[W,K^{(\vec \beta, \gamma)}]]\right)\right]\delta t + \mathcal{O}(\delta t^2).
\end{eqnarray}
Thus, if we choose $h_\gamma = [T,G^{(\vec \beta, \gamma)}] - 2\mathrm{tr}_2 [[W,K^{(\vec \beta, \gamma)}]] = - h_\gamma^T$ we have defined an evolution with an orthogonal matrix that decreases the energy and, for large $t$, results in a state that fulfills the steady-state condition Eq.(\ref{eq:cond_energy}). Furthermore, the calculation of $h_{\gamma}$ requires the summation over $O(N^3)$ paramters and can thus be performed efficiently.

\section*{Appendix D: Pairing}

In the following we present a possibility to calculate numerically the pairing of a state $|\phi_{\vec \beta, \gamma}\rangle$ with $2N$ particles and $4M$ modes via
\be
\label{eq:calc_pairing}
\mathcal{M}(\rho)  = \frac{1}{N}\max_{\{\ad_i\}_i}\sum_{k,l=1}^{2M}|\langle \ad_{2k-1} \ad_{2k}a_{2l}a_{2l-1}\rangle_{\rho}|.
\ee
We perform the optimization over all possible set of bases in the following way: First, we choose a fixed basis $\mathcal{B}_0$ by making the identification $a_{2k-1} \leftrightarrow a_{x\uparrow}$ $a_{2k} \leftrightarrow a_{x\downarrow}$. The set of operations that are linking two different sets of modes are called passive transformations. They can be represented by a unitary operator $U_{H_p} = e^{iH_p}$, where $H_p = \sum_{k,l}h_{k,l}\ad_ka_l$ with $h^{\dagger} = h$. Thus, we can write the optimization in Eq.~(\ref{eq:calc_pairing}) in the following way: Define the operator $P$ that calculates the pairing in the basis $\mathcal{B}_0$,
\be
P = -\sum_{x,y \in \Lambda} \ad_{x\uparrow} \ad_{x\downarrow}a_{y \downarrow}a_{y \uparrow} = i\sum_{kl}T^{(P)}_{kl}c_kc_l + \sum_{klmn}W^{(P)}_{klmn}c_kc_lc_mc_n.
\ee
Then, 
$$ \mathcal{M}(\rho)  = \frac{1}{N}\max_{U_{H_P}}\sum_{x,y, \in \Lambda}\langle \phi_{\vec \beta, \gamma}|U_{H_P}^\dagger \ad_{x \uparrow} \ad_{x \downarrow}a_{y \downarrow}a_{y \uparrow} U_{H_P}|\phi_{\vec \beta, \gamma}\rangle = \frac{1}{N} \langle \phi_{\vec \beta, \gamma}|U_{H_P}^\dagger P U_{H_P}|\phi_{\vec \beta, \gamma}\rangle.$$
Note that we could get rid of the absolute values since we can always find a passive transformation so that all the terms in the sum of Eq.~(\ref{eq:calc_pairing}) are positive. Now, in order to be able to use techniques we have developed for calculating the ground state energy in Appendix C, we define the set $\mathcal{S}_{|\phi_{\vec \beta, \gamma}\rangle}$ of all states that can be brained from the state $|\phi_{\vec \beta, \gamma}\rangle$ via a passive transformation. Then, it follows immediately that
\begin{align}
- N \mathcal{M}(|\phi_{\vec \beta, \gamma}\rangle)  &=\min_{|\Psi\rangle \in \mathcal{S}_{|\phi_{\vec \beta, \gamma}\rangle}} \langle \Psi|P |\Psi\rangle =\min_{U_{H_P}} \langle \phi_{\vec \beta, \gamma} | U_{H_p}^\dagger P U_{H_p} |\phi_{\vec \beta, \gamma}\rangle\nonumber\\
& = \min_{U_{H_P}}\left(  -\mathrm{tr}[G^{(\vec \beta, \gamma)}O_{H_p}^TT^{(P)}O_{H_p}]+ \mathrm{tr}[K^{(\vec \beta, \gamma)} (O_{H_p}^T\otimes O_{H_p}^T)W^{(P)}(O_{H_p}\otimes O_{H_p})]\right)=  \min_{U_{H_P}} \mathcal{P}_{\vec \beta, \gamma}(H_p),
\end{align}
where $O_{H_p}$ is the orthogonal matrix realizing the passive transformation $U_{H_p}$ on the Majorana operators. Now, we follow a similar route as in the case of the minimization of the energy explained in Appendix C: We introduce a time dependence in the Hamiltonian $H_p$ and write $O_{H_p(t+\delta t )} = e^{h_{H_p}\delta t}O_{H_p(t)}$. Then,
\begin{eqnarray}\label{eq:maximize_pairing}
&\mathcal{P}_{\vec \beta, \gamma}(H_p(t+\delta t)) - \mathcal{P}_{\vec \beta, \gamma}(H_p(t)) =\mathrm{tr}\left[h_{H_p}\left([T^{(P)},G^{(\vec \beta, \gamma)}] - 2\mathrm{tr}_2 [[W^{(P)},K^{(\vec \beta, \gamma)}]]\right)\right]\delta t = 2 \mathrm{tr}[h_{H_p}Z],
\end{eqnarray}
where we introduce a block decomposition for the matrix $Z \in \mathbb{R}^{8N \times 8N}$
\be
Z =[T^{(P)},G^{(\vec \beta, \gamma)}] - 2\mathrm{tr}_2 [[W^{(P)},K^{(\vec \beta, \gamma)}]] = \left(\begin{array}{cc}
Z_{11} & Z_{12} \\
-Z_{12}^T & Z_{22}
\end{array}\right),
\ee
where $Z_{i,j} \in \mathbb{R}^{4N \times 4N}$. Now, in contrast to Appendix C, we cannot choose $h_{H_p} = Z$, since $U_{H_p}$ is a  passive transformation  which imposes further constraints on $h_{H_p}$: If we rewrite the Hamiltonian $H_p$ in the basis of Majorana operators, we arrive at
\be
H_p = \vec c^{\,\,T}
\left(\begin{array}{cc}
h_I & -h_R \\
h_R & h_I
\end{array}\right)\vec c = \vec c^{\,\,T}  h_c \vec c.
\ee
 Here, the $2N \times 2N$ matrices  $h_R = {\rm Re}(h) = h_R^T$ and $h_I = {\rm Im}(h) = -h_I^T$ are the real and imaginary part of $h_{H_p} \in \mathbb{C}^{4N \times 4N}$. Thus, $h_{H_p}$ has to be of the same block form as $h_c$. Then, it is easy to check that by choosing
\begin{eqnarray}
 h_I &= Z_{11} + Z_{22}\\
 h_R &= -(Z_{12}+Z_{12}^T)
 \end{eqnarray}
the pairing decreases.

\section{References}

\end{document}